\documentclass[preprint,preprintnumbers,aps,nofootinbib,tightenlines,floatfix,showpacs,superscriptaddress]{revtex4}
\usepackage{amssymb,latexsym}
\usepackage{amsmath,amsbsy,bbm}
\usepackage{epsfig,bm}
\usepackage{graphicx,comment}
\usepackage{bm,comment}
\usepackage{epstopdf}
\usepackage{color}
\usepackage{subfigure}
\usepackage{hyperref}

\DeclareMathOperator{\tr}{tr}

\def\Dslash{D\hskip-0.65em /}

\def\tr{\text{tr}}

\def\cS{{\mathcal S}}

\def\comma{{\; ,}}
\def\Nt{{N_{\tau}}}
\def\Ns{{N_{\sigma}}}
\def\ass{{a \sqrt{\sigma}}}
\def\eqref#1{{(\ref{#1})}}

\def\Dslash{D\!\!\!\!/}

\newcommand\contains{\ensuremath{\ni}}

\newcommand{\Real}{\ensuremath{\mathbb{R}}}
\newcommand{\Rounded}{\ensuremath{\mathbb{Z}}}
\newcommand{\Artifact}{\ensuremath{a}}
\newcommand{\Fit}{\ensuremath{f}}

\newcommand{\best}[1]{\textbf{#1}}

\begin{document}

\preprint{LLNL-JRNL-671143, INT-PUB-15-017}

\title{Lattice QCD input for axion cosmology}

\author{Evan Berkowitz}
\affiliation{%
Physical and Life Sciences Directorate,
Lawrence Livermore National Laboratory,
Livermore, California 94550, USA}

\author{Michael I. Buchoff}
\affiliation{%
Institute for Nuclear Theory, Box 351550, Seattle, WA 98195-1550, USA}

\author{Enrico Rinaldi}
\affiliation{%
Physical and Life Sciences Directorate,
Lawrence Livermore National Laboratory,
Livermore, California 94550, USA}


\pacs{12.38.Gc, 14.80.Va}

\begin{abstract}
One intriguing beyond-the-Standard-Model particle is the QCD axion, which could simultaneously provide a solution to the Strong CP problem and account for some, if not all, of the dark matter density in the universe.
This particle is a pseudo-Nambu--Goldstone boson of the conjectured Peccei--Quinn (PQ) symmetry of the Standard Model. Its mass and interactions are suppressed by a heavy symmetry breaking scale, $f_a$, whose value is roughly greater than $10^{9}$ GeV (or, conversely, the axion mass, $m_a$, is roughly less than $10^4\ \mu \text{eV}$).
The density of axions in the universe, which cannot exceed the relic dark matter density and is a quantity of great interest in axion experiments like ADMX, is a result of the early-universe interplay between cosmological evolution and the axion mass as a function of temperature.
The latter quantity is proportional to the second derivative of the temperature-dependent QCD free energy with respect to the CP-violating phase, $\theta$. However, this quantity is generically non-perturbative and previous calculations have only employed instanton models at the high temperatures of interest (roughly 1 GeV).
In this and future works, we aim to calculate the temperature-dependent axion mass at small $\theta$ from first-principle lattice calculations, with controlled statistical and systematic errors.
Once calculated, this temperature-dependent axion mass is input for the classical evolution equations of the axion density of the universe, which is required to be less than or equal to the dark matter density.  Due to a variety of lattice systematic effects at the very high temperatures required, we perform a calculation of the leading small-$\theta$ cumulant of the theta vacua on large volume lattices for SU(3) Yang--Mills with high statistics as a first proof of concept, before attempting a full QCD calculation in the future.  From these pure glue results, the misalignment mechanism yields the axion mass bound $m_a \geq (14.6\pm0.1) \ \mu \text{eV}$ when PQ-breaking occurs after inflation.

\end{abstract}
\maketitle

\section{Introduction}  \label{sec:intro}
Despite overwhelming experimental and theoretical evidence that Quantum Chromodynamics (QCD) is the underlying theory of interactions between quarks and gluons, one puzzle has eluded explanation for over 30 years: the Strong CP problem~\cite{tHooft:1976fv,Callan:1976je,Jackiw:1976pf}.
QCD allows for a $\mathcal{O}(1)$ combined charge-conjugation and parity (CP) violating phase $\theta$ and, yet, this phase is found experimentally to be consistent with zero to within one part in ten billion~\cite{Baker:2006ts}.

Other than attributing this result to an appreciable fine-tuning, there have been three proposed explanations explored at length: the possibility of a zero up-quark mass~\cite{Kaplan:1986ru}, spontaneous CP breaking~\cite{Beg:1978mt,Mohapatra:1978fy,Georgi:1978xz,Nelson:1983zb,Barr:1984qx,Nelson:1984hg,Vecchi:2014hpa}, and an additional ``hidden'' chiral symmetry.
After taking into account the up-to-date Standard Model flavor-changing constraints~\cite{Agashe:2014kda} and lattice QCD calculations~\cite{Aoki:2013ldr}, the most feasible proposal is the last.
The additional chiral symmetry proposed, also referred to as Peccei--Quinn (PQ) symmetry~\cite{Peccei:1977hh,Peccei:1977ur}, is such that upon spontaneous symmetry breaking, the effective potential naturally gives a zero CP-violating phase\footnote{Even this explanation for a small $\theta$ has limitations, since PQ-violating Plank-scale operators up to dimension 10 can reintroduce the Strong CP problem~\cite{Holman:1992us,Cheung:2010hk}.}.
As with any spontaneous breaking of a continuous symmetry, one would expect a resulting (pseudo-)Nambu--Goldstone particle to be present in the universe; this particle is known as the axion\footnote{For some general reviews on the subject of axions, see Refs.~\cite{Peccei:2006as,Kim:2008hd}.}~\cite{Peccei:1977hh,Peccei:1977ur,Weinberg:1977ma, Wilczek:1977pj}.

The original proposals focused on axions masses at or below the electroweak scale~\cite{Peccei:1977hh,Peccei:1977ur}, but collider~\cite{Bardeen:1986yb,Asano:1981nh} and astrophysical constraints~\cite{Turner:1989vc,Raffelt:1990yz,Raffelt:1999tx} now require the axion to have a mass below $10^4\ \mu \text{eV}$.
A suitable class of axion models, named ``invisible axions'', could allow for light axions whose interactions are suppressed by very high energy scales~\cite{Dine:1981rt,Kim:1979if,Shifman:1979if, Zhitnitsky:1980tq}.
In these models the large density of light axions could potentially account for some, if not all, of the dark matter density in the universe~\cite{Preskill:1982cy,Abbott:1982af,Dine:1982ah}.
For this reason, dedicated experiments such as ADMX~\cite{Sikivie:1983ip, Asztalos:2009yp, Asztalos:2011ei} search for the axion coupling to two photons directly.
With the next generation of axion experiments underway~\cite{Asztalos:2011ei} along with increased constraints on inflation~\cite{Ade:2015lrj}, a great deal is expected to be learned about potential axion interactions in the next few years.

There has been a wealth of research on axion cosmology and the related axion energy density over the past 30 years~\cite{Preskill:1982cy,Abbott:1982af,Dine:1982ah,Davis:1985pt, Davis:1986xc, Harari:1987ht, Davis:1989nj,Hagmann:1990tj,Nagasawa:1994qu,Chang:1998tb,Hagmann:1998me, Yamaguchi:1998gx, Yamaguchi:1999dy, Hagmann:2000ja, Dabholkar:1989ju, Battye:1994au, Sikivie:2006ni,Bae:2008ue, Wantz:2009it, Hiramatsu:2012gg, Kawasaki:2013ae}. 
One particular constraint, the ``overclosure bound'', requires that the axion density today not be greater than the total observed present-day dark matter abundance.
This bound requires accounting for the evolution of the axion mass through cosmological history and its consequences for axion production.
Depending on when PQ~breaking occurs relative to inflation, the overclosure bound gives different relations.  If PQ~breaking occurs during or before inflation, the bound is on a relation of two variables, the initial value for the CP-violating phase inside our cosmic horizon and the axion decay constant, $f_a$.  If PQ~breaking occurs after inflation, the bound is solely on $f_a$.  After inflation, there are three stages to this early universe, high-temperature evolution: the post-inflation evolution to a time when the axion mass is comparable to the inverse horizon of the universe (i.e. the Hubble constant), the subsequent evolution in the chirally symmetric phase, and the time period before and after the chiral symmetry breaking phase transition of QCD.
In the first stage, the axion mass depends greatly on the QCD free energy which is a function of the CP-violating phase, $\theta$, and temperature, $T$.
To date, the only methods employed for estimating this QCD free energy dependence as related to the axion mass are the dilute instanton gas model (DIGM)~\cite{Gross:1980br} and the interacting instanton liquid model (IILM)~\cite{Schafer:1996wv}. The DIGM is expected to be valid only for high temperatures, while the IILM models strong interactions between instantons around the QCD phase transition.
In the context of axions, both of these scenarios have been explored in detail~\cite{Turner:1985si, Wantz:2009it}.
While the desired characteristic temperatures are indeed large ($\sim1$ GeV), one open question is to what extent the DIGM and IILM are valid (i.e. to what extent non-perturbative physics plays a role at these temperatures and is modeled properly) and how accurate the overall scale is (the free energy scales as the confinement scale, $\Lambda_{QCD}$, to a large power).
 Another related issue is that neither the DIGM nor the IILM yield controlled uncertainties from first principles.
The only known approach that can address these two quandaries\footnote{For these reasons, the use of lattice QCD was suggested in Ref.~\cite{Wantz:2009it}.} quantitatively is lattice gauge theory, which is what we apply in this work.

There have been ample lattice studies of QCD vacuum properties at high temperature, highlighted by recent work verifying the QCD critical temperature ($T_c$) with 2+1 quarks at physical pion masses in a chiral discretization~\cite{Bhattacharya:2014ara}, and controlled continuum limits for the QCD equation of state~\cite{Borsanyi:2013bia,Bazavov:2014pvz} with temperatures as large as $T\sim 600 \ \text{MeV}$.
Unfortunately, simulating higher temperatures becomes computationally more expensive due to unphysical systematic finite-volume effects (for a fixed number of lattice sites, the physical volume decreases as the temperature increases).
However, simulations of pure Yang--Mills with three colors is appreciably cheaper and high-temperature $\theta=0$ studies spanning $T \sim (5 - 1000) T_c$ suggest that non-perturbative effects can still be appreciable at $T \sim 1 \ \text{GeV}$~\cite{Borsanyi:2012ve}.
Appreciably more difficulty ensues when exploring $\theta \neq 0$ quantities, as the topological term in the Lagrangian has an associated sign problem that renders standard lattice Monte-Carlo techniques intractable\footnote{For progress in solving sign problem for actions with non-zero $\theta$, see Ref.~\cite{Bongiovanni:2014rna,Aarts:2014kja}.}.
However, techniques do exist for extracting quantities at small $\theta$, such as the leading $\theta^2$ dependence of the free energy which is proportional to the topological susceptibility\footnote{For a review on lattice calculations of topological susceptibilities, see Ref.~\cite{Vicari:2008jw}.}.
While topological fluctuations are a topic of active focus for lattice calculations at zero-temperature~\cite{Gattringer:2001ia,DelDebbio:2002xa,Durr:2006ky,Cichy:2014qta}, there have only been a handful of studies at finite temperature; first for Yang--Mills~\cite{Alles1996,Gattringer:2002mr,DelDebbio:2004rw,Cossu:2013uua} and more recently in full QCD using chiral discretizations~\cite{Buchoff:2013nra}.

In this work, we aim to extend and improve the finite-temperature results in Ref.~\cite{Gattringer:2002mr} with the express purpose of comparing with the DIGM and IILM and, within the capability of this calculation, quote a first-principles bound for the axion mass with controlled uncertainties.
It is important to note that this work is only an initial step towards the full QCD problem and it contains two primary inadequacies due to limited resources and lattice technology.

The first inadequacy is that our calculations are performed for a 3-color pure Yang--Mills theory without the dynamical fermions of QCD.
Ultimately, this may not prove to be too far from the physical result, as effects of fermion loops are expected to be suppressed at high temperatures~\cite{Braaten:1995cm}.  For topological quantities, however, this may not be the case~\cite{Kanazawa:2014cua} and full QCD calculations should be pursued in the future.
The reasoning for studying a purely gluonic theory at this stage is two-fold.
First, Yang--Mills theories are over ten times cheaper to simulate than full QCD due to the ability to employ the heatbath algorithm for entirely bosonic theories.
This extra gain in computational efficiency will allow us to go to higher temperatures and larger volumes, and to directly estimate systematic effects for topological quantities.
The second advantage is that heatbath Monte-Carlo calculations have appreciably shorter autocorrelations in extracting topological quantities than the hybrid Monte-Carlo algorithm used for dynamical fermions.
This will allow us to extract the high level of independent statistics required to address the problem at hand.

The second inadequacy in our present calculation is that reaching the large values of $T/T_c \sim 5$ with controlled volume systematics is still beyond our current computational resource limitations\footnote{Future explorations of extracting topological quantities on anisotropic lattices may alleviate some of the computational burden on volume as long as discretization effects stay minimal.}.
Summarizing our calculation, we aim to test the validity and overall scale of the DIGM/IILM for a SU(3) Yang--Mills theory for $T\sim 372 - 710$ MeV (or, alternatively, $T/T_c \sim 1.31 - 2.50$, remembering that $T_c$ for SU(3) Yang--Mills is $\approx 284$ MeV~\cite{Lucini:2005vg}) with high statistics and controlled volume and discretization uncertainties.
From these results, we extrapolate our results to extract the characteristic temperature when the axion mass and Hubble constant are comparable and proceed to evaluate what this result would imply for the axion mass in the present day universe with propagated uncertainties.

The next section is a brief review of the cosmological evolution of axions, while Sec.~\ref{sec:energy} describes the free energy of QCD in the presence of a $\theta$ term and a finite temperature.
The paper then continues with a detailed description of the lattice simulations in Sec.~\ref{sec:lattice}, lattice results in Sec.~\ref{sec:analysis} and lattice error budget in Sec.~\ref{sec:syst}. Based on first-principles lattice results we derive a bound on the axion mass from the aforementioned overclosure argument in Sec.~\ref{sec:res} and Sec.~\ref{sec:bounds} before our concluding remarks.

\section{Brief review of axion cosmology} \label{sec:cosmo}
The theory of quarks and gluons, Quantum Chromodynamics (QCD), supports a term in its action $\mathcal{S}_{\text{QCD}}$ which violates the combined charge-conjugation and parity symmetry (CP),
\begin{align}\label{eq:theta_term_topology}
	\mathcal{S_{\text{QCD}}} &\contains \theta Q &	Q &= \int d^4x\;\frac{g^2}{32\pi^2} \tr{F\tilde{F}}
\end{align}
where $Q$ is the topological charge, $F$ ($\tilde{F}$) is the (dual) gauge field strength tensor, and $g$ the QCD gauge coupling constant.
The Strong CP problem is the observation that the parameter $\theta$ could in principle take any value between the maximally-CP-violating values of $-\pi$ and $\pi$, but is measured to be consistent with the CP-conserving value of zero to one part in ten billion~\cite{Baker:2006ts}.
An elegant explanation of the small value of $\theta$ was proposed by Peccei and Quinn: it is possible to promote $\theta$ to a dynamical variable in such a way that it is controlled by a (pseudo-)Nambu--Goldstone boson called the axion.
The Peccei--Quinn (PQ) symmetry makes the axion naturally light and creates a potential for $\theta$ which favors small $\theta$.
The QCD+axion Lagrangian is given by
\begin{equation}
\mathcal{L}=-\frac{1}{4}F^a_{\mu\nu}F^{a\mu\nu}+\frac{1}{2}\partial_\mu a \partial^\mu a +\sum_q \bar{q}(i \Dslash -m_q)q +\frac{g^2}{32\pi^2}\left(\theta + \frac{a}{f_a}\right) F^a_{\mu\nu}\tilde{F}^{a\mu\nu},
\end{equation}
where $f_a$ is proportional to the vacuum expectation value that breaks PQ~symmetry.
This quantity is the one free parameter of standard QCD axion theories, and it is this scale which ultimately determines the axion mass and its couplings to two photons that experiments are actively searching for~\cite{Sikivie:1983ip, Asztalos:2009yp, Asztalos:2011ei}.

While $f_a$ is the primary parameter for both the axion mass and axion energy density given the QCD Lagrangian, non-perturbative QCD effects lead to non-trivial cosmological consequences, especially for the axion number density.
At low temperatures, after chiral symmetry has been broken, the axion mass $m_a$ and coupling obeys a relatively simple relation~\cite{Weinberg:1977ma,Dine:1982ah, Abbott:1982af,Preskill:1982cy}
\begin{equation}\label{eq:axion_mass_today}
m_a f_a = \frac{\sqrt{m_u m_d}}{m_u+m_d} f_\pi m_\pi,
\end{equation}
where $m_\pi$ and $f_\pi$ are the pion mass and pion decay constant, respectively.   The axion mass at temperatures in the chirally-symmetric phase of QCD will be the primary focus of this work.

One significant constraint on axions is that their present energy density does not exceed the dark matter density of the universe (the ``overclosure'' bound).
To determine the relevant energy density today, one must explore the interplay between the axion mass and cosmological evolution by solving the equations of motion.
However, the initial question in this evolution is to place the moment PQ~symmetry breaks along the cosmic timeline.
There are two options:
\begin{enumerate}
	\item PQ~symmetry breaks before or during inflation. 
	\item PQ~symmetry breaks after inflation.
\end{enumerate}

While the final energy density comes down to one (or effectively two) free parameters, there are many non-trivial calculations and subtleties that need to be understood from both cosmology and non-perturbative field theory.
It is useful to give a brief summary on how the energy density arises in each of these PQ-breaking scenarios\footnote{For detailed reviews, axion cosmology, see Ref.~\cite{Sikivie:2006ni,Wantz:2009it,Kawasaki:2013ae}}:
\begin{enumerate}
\item For PQ~symmetry breaking before or during inflation, the field $\theta=a/f_a$ would be homogenized over large distances.
  In other words, the causally disconnected regions of spacetime that have different initial values of the field (angle) before inflation stretch to cosmic scales, so that our universe has a uniform initial $\theta$.
  Any excited axion mode or topological defect (e.g. strings) will be diluted away, leaving only the zero-mode contributions to the energy density.
  The resulting axion density arises from this ``misalignment mechanism'' and has large dependence on this initial theta angle (often referred to as the misalignment angle, $\theta_1$, which can take any value between $0$ and $2\pi$).
  Ultimately, the final energy density will be proportional to roughly $\theta_1^2$, which is effectively a free parameter along with $f_a$.
  Thus, the overclosure bound can only bound $f_a$ as a function of $\theta_1$.
  Also, due to large fluctuations during inflation, isocurvature density perturbation bounds from CMB observations can put important constraints on $\theta_1$ as a function of the Hubble scale during inflation~\cite{Axenides:1983hj, Seckel:1985tj, Linde:1987bx,Linde:1990yj, Linde:1991km, Lyth:1991ub,Kawasaki:2013ae}.
\item For PQ~symmetry breaking after inflation, the misalignment angle $\theta_1$ is (effectively) averaged over, since all values $\in [0,2\pi]$ are equally probable in small regions of our universe. 
  Moreover, there are several additional effects that must be accounted for in addition to the misalignment mechanism in this case.
  First, nonzero-momentum modes can contribute and their effect is discussed in Refs.~\cite{Chang:1998tb,Sikivie:2006ni}.
  Second, when PQ~symmetry is broken, global topological defects called axionic strings will form and decay to axions, introducing an efficient process for energy loss and effectively increasing the total axion number density.
  It is currently debated as to how much these decays will increase the total axion number, with some claiming between one and two orders of magnitude more than the misalignment mechanism~\cite{Davis:1986xc,Davis:1989nj,Dabholkar:1989ju,Yamaguchi:1998gx,Hiramatsu:2010yu}, while other claim it is on the same order~\cite{Harari:1987ht,Hagmann:1990mj, Hagmann:2000ja}.
  The third effect is the axion strings can be connected to one or more domain walls~\cite{Sikivie:1982qv} which correspond to different minima of the axion potential.
  The decay of these domain walls also leads to additional axions, but is believed to be subdominant to string decay ~\cite{Nagasawa:1994qu, Chang:1998tb, Hagmann:1998me, Hiramatsu:2012gg}.
  Overall, the energy density can be calculated in terms of one free parameter, $f_a$ and, as a result, an overclosure bound on $f_a$ can be derived.
\end{enumerate}

Our goal in this and future work is to improve the non-perturbative QCD input that goes into the axion density calculations.
In particular, we aim to provide a controlled calculation of the temperature-dependent axion mass from first-principle lattice calculations\footnote{Other non-perturbative QCD aspects include the number of theta-vacua/domain walls for $\theta$ changing by $2\pi$~\cite{Sikivie:1982qv} and whether or not axion number density is constant throughout the QCD phase transition~\cite{DeGrand:1985uq,Bae:2008ue}.
We do not discuss these further in this work.}.
The axion mass in the chirally symmetric phase is given by
\begin{equation}
m_a^2(T) f_a^2= \frac{\partial^2 F(\theta,T)}{\partial \theta^2}\bigg|_{\theta=0}\equiv \chi(T),
\end{equation}
where $F(\theta,T)$ is the QCD free energy as a function of CP-violating phase and temperature, while $\chi$ is the topological susceptibility.
In each of the energy density scenarios discussed above, the temperature-dependent axion mass $m_a(T)$ plays a role, particularly when this quantity is comparable to the Hubble scale in the early universe evolution.
We will restrict our discussion primarily to the analytic evaluations of the misalignment mechanism, as the lattice calculations performed in this work could be appreciably different from QCD due the computational and algorithmic limitations discussed in subsequent sections.
While we will not discuss it in detail, we will also point out how the axion mass enters into the relevant calculations for axion strings and domain walls.
Once the full QCD results are at a mature stage, the lattice result for  $m_a(T)$ should be used as input for the numerical solutions to the classical equations of motions and cosmology simulations.

\subsection{Misalignment Mechanism}\label{sec:misalignment}
To keep the presentation self-consistent, we summarize the description of the misalignment mechanism of Ref.~\cite{Preskill:1982cy} and subsequently summarized in Ref.~\cite{Sikivie:2006ni,  Wantz:2009it, Kawasaki:2013ae}.
We start from  from the Robertson--Walker metric of the universe,
\begin{equation}
-ds^2 = dt^2 - R(t)^2 d\mathbf{x} \cdot d\mathbf{x} \ ,
\end{equation}
and write down the axion equation of motion:
\begin{equation}\label{eq:EOM}
\left(\partial_t^2+3H\partial_t - \frac{1}{R^2}\nabla_x^2\right)a(x)+m_a^2(t)f_a\sin\left(\frac{a(x)}{f_a}\right)=0 \ ,
\end{equation}
where $H=\dot{R}/R$ is the Hubble constant and the second term is given by the derivative of the effective potential for the axion field, $V_a$, given by,
\begin{equation}
V_a = f_a^2m_a^2(t)\left[1-\cos\left(\frac{a}{f_a}\right)\right] \ .
\end{equation}
Along with the axion equation of motion, the Hubble constant evolution is given by the Einstein equation 
\begin{equation}\label{eq:Hubble}
H^2 = \frac{1}{3M_{pl}^2}\left\{\frac{\pi^2}{30}g_{*,R}T^4 + f_a^2\left(\frac12 \left(\frac{da}{dt}\right)^2+m_a^2(t)\left[1-\cos\left(\frac{a}{f_a}\right)\right]\right)\right\} \ ,
\end{equation}
where $M_{pl}$ is the reduced Planck mass $\sqrt{\hbar c / 8 \pi G}$.
For given $m_a(t)$, or equivalently $m_a(T)$, one can solve Eq.~\eqref{eq:EOM} and Eq.~\eqref{eq:Hubble} numerically to arrive at the axion energy density:
 \begin{equation}
\rho_a = \frac{1}{2}\left(\frac{da}{dt}\right)^2 + \frac{1}{2}m_a^2(t)a^2(x) \ . 
\end{equation}
At this point, it is useful to discuss the qualitative consequences of these solutions (see Fig.~4 in Ref.~\cite{Wantz:2009it} for an illuminating plot).
When considering early times, one notes that the Hubble constant is much larger than the axion mass ($3H\gg m_a$), meaning that the axion wavelength is larger than the Hubble length.
Thus, the axion does not feel a potential and it is effectively massless.
Moreover, the axion number density is zero and the axion field has a constant value, $\theta_1$ (the misalignment angle).
As time increases (and temperature decreases), the axion mass increases while the Hubble constant decreases.
The solutions change drastically when the axion mass is of the same order as the Hubble constant ($3H\approx m_a$), at which point the axion mass ``turns on'' and the axion field rolls down the potential---during this period the axion number density jumps to a nonzero value.
From this point in time onward, both the axion field and axion number density have decaying oscillations about their eventual final values that we see today (the decay becomes adiabatic when $3H\ll m_a$).    

Analytic progress can be made by making a few key observations\footnote{For simplicity, we look at just the zero-momentum mode, dropping the $\nabla_x^2$ from Eq.~\eqref{eq:EOM}.
For a complete calculation with the non-zero modes, see Ref.~\cite{Chang:1998tb,Sikivie:2006ni}.}.
First, let us quantify the characteristic temperature and corresponding time, $T_1$ and $t_1$, respectively, when the Hubble constant is comparable to the axion mass
\begin{equation}\label{eq:ma_H}
  m_a(T_1) = 3H(T_1) \equiv \frac{1}{t_1} \ .
\end{equation}
When $T\gtrsim T_1$,  Eq.~\eqref{eq:Hubble} reduces to
\begin{equation}\label{eq:H1Gev}
  H(T) = \frac{\pi g_*^{1/2}(T)T^2}{\sqrt{90}M_{Pl}}
\end{equation}
where $g_*(T)$ is the effective number of relativistic degrees of freedom at a given temperature.  In this work, we use the parameterization of $g_*(T)$ in Appendix A of Ref.~\cite{Wantz:2009it}.
We will compare our lattice $m_a(T)$ to $H(T)$ using Eq.~\eqref{eq:ma_H} to extract $T_1$.
The second observation is that the decaying oscillations when $3H\ll m_a$ are adiabatic and the equation of motion can be recast as
\begin{equation}
  \frac{1}{f_a^2}\frac{d\rho_a}{dt}=2m_a(t)\frac{d m_a}{dt}\left[1-\cos\left(\frac{a}{f_a}\right)\right]-3H\frac{da}{dt} \ .
\end{equation}
When $m_a\gg H$ and $m_a\gg dm_a/dt$, this relation leads to the adiabatic invariant
\begin{equation}\label{eq:Inv}
  \frac{\rho_a R^3}{m_a}  = \text{constant} \ ,
\end{equation}
and subsequently~\cite{Wantz:2009it},
\begin{equation}
  \rho_a(T_\gamma) = \rho_a(T_1) \frac{m_a(T_\gamma)}{m_a(T_1)}\left(\frac{R(T_1)}{R(T_\gamma)}\right)^3\simeq \frac{\rho_a(T_1)}{\gamma} \frac{m_a(T_\gamma)}{m_a(T_1)}\frac{g_*(T_\gamma) T_\gamma^3}{g_*(T_1)T_1^3} \ ,
\end{equation}
where $s$ is the entropy density, $T_\gamma \simeq 2.73 \ \text{K}$ is the temperature of the cosmic microwave background today and $\gamma$ is the ratio of the entropy density today to the entropy density at $t_1$.
From this relation, along with the fact that $\rho_a(T_1) \sim \frac{1}{2}m_a(T_1)^2  f_a^2 \theta_1^2$ and Eq.~\eqref{eq:axion_mass_today}, Eq.~\eqref{eq:ma_H}, and Eq.~\eqref{eq:H1Gev}, we arrive at the commonly used relation~\cite{Bae:2008ue}
\begin{equation}\label{eq:density}
  \rho_a(T_\gamma) = \left[\frac{3\pi g_*(T_\gamma)}{\sqrt{90g_*(T_1)}}\frac{\sqrt{m_u m_d}}{m_u + m_d}\frac{f_\pi m_\pi T_\gamma^3}{M_{Pl}}\right]\left(\frac{f_a}{T_1}\right)\left(\frac{F_1(\theta_1)}{2\gamma}  \right) \theta_1^2 \ ,
\end{equation}
where $F_1(\theta_1)$ accounts for the anharmonic corrections to Eq.~\eqref{eq:Inv}.
If it is also assumed that the expansion of the universe is adiabatic for temperatures below $T_1$, the quantity $F_1(\theta_1)/2 \gamma \approx 1$ for $\theta_1 \lesssim 2$~\cite{Bae:2008ue}.
Using lattice QCD calculations, $T_1$ can be extracted as a function of $f_a$ and a confinement-scale observable (such as the QCD deconfinement temperature $T_c$), and as a result, a bound on the free parameters $f_a$ and $\theta_1$ can be derived to ensure that the axion density is below that of the dark matter density.
If we are only exploring axion theories where PQ-breaking occurs after inflation, the misalignment angles are averaged over, and the $\theta_1$ dependence in Eq.~\eqref{eq:density} is replaced with the average value, $\theta_1^2 \rightarrow\langle \theta_1^2 \rangle = \pi/\sqrt{3}$.

\subsection{Axion Strings}
In the case where PQ~breaking occurs before or during inflation, any resulting topological defects, such as axion strings or domain walls, are diluted away and only the misalignment mechanism contributes to the axion density with the fixed $\theta_1$ of our observed universe.  However, if PQ~breaking occurs after inflation, these topological defects can decay into axions and significantly alter the number density of axions in the universe today. Scenarios with two or more decaying domain walls are highly constrained from neutron EDM bounds on CP violation and would require an appreciable fine-tuning~\cite{Kawasaki:2013ae}.  

The current understanding is that the quantity of axions produced by domain wall decay is below the number of axions produced from string decay, a number which is at least comparable to the density produced from the misalignment mechanism and could even be appreciably larger as debated in the literature~\cite{Davis:1986xc,Davis:1989nj,Dabholkar:1989ju,Yamaguchi:1998gx,Hiramatsu:2010yu,Harari:1987ht,Hagmann:1990mj, Hagmann:2000ja}.  In the case of string decay, this process occurs for string frequencies between the Hubble scale and the axion mass ($H< \omega < m_a$), and thus begins to play a role at temperature $T_1$.  Similarly, domain walls also have a dependence on $m_a$, which plays a role in the axion spectrum.  While it is not generally expected to be the primary source of uncertainty in these large scale cosmological simulations~\cite{Wantz:2009it,Hiramatsu:2010yu}, accurate input for the axion mass as a function of temperature would nonetheless be useful.

\section{QCD free energy as a function of $T$ and $\theta$}\label{sec:energy}
The free energy of QCD, $F$, as a function of temperature and theta is given in terms of the path integral
\begin{equation}
Z_{QCD}(\theta, T) = \int [dA] [d\psi] [d\bar{\psi}] \exp\left(-T\sum_t d^3 x\; \mathcal{L}_{QCD}(\theta)\right) = \exp [-VF(\theta,T)],
\end{equation}
where 
\begin{equation}
\mathcal{L}_{QCD}(\theta) = \mathcal{L}_{QCD} + \frac{g^2\theta}{32\pi^2}\epsilon^{\mu\nu\rho\sigma}F^a_{\mu\nu}F^{a}_{\rho\sigma}.
\end{equation}
The free energy is periodic in $\theta$ and thus we can restrict to the range $-\pi <\theta <\pi$ and parameterize the free energy in terms of a sum of cosine functions,
\begin{equation}
F(\theta,T) = \sum_n C_n(T) \cos(n\theta).
\end{equation}
Since QCD is a non-perturbative problem for generic $T$ and $\theta$, the values of $C_n(T)$ are not readily accessible; even in first-principle lattice QCD calculations, non-zero $\theta$ introduces a computationally intractable sign problem.
However, at high enough temperatures, QCD interactions are perturbative and their parametric dependence on $T$ and $\theta$ should be given by the dilute instanton gas model (DIGM)~\cite{Gross:1980br}.
In this model, where the QCD background is approximated by non-interacting instantons, only $C_1$ contributes to the free energy (all $n>1$ coefficients are zero).
The DIGM also predicts the value of the coefficient (whose mass dimension is 4) as a function of $\Lambda_{QCD}$,
\begin{equation}
F_{DIGM}(\theta,T) = -C(T) \cos(\theta),
\end{equation}
where $C(T)$ can be well-approximated by~\cite{Turner:1985si,Bae:2008ue}
\begin{equation}
C(T) \simeq \frac{C\Lambda^4}{(T/\Lambda)^n},
\end{equation}
where $C$ is a dimensionless constant in temperature and the fermion masses and $n$ is a power that is a function of number of flavors and colors\footnote{The actual expression derived from DIGM contains a more involved integral over the logarithmic running of the coupling.
For a complete expression see Appendix A of Ref.~\cite{Turner:1985si}}.
The $\Lambda$ in this equation essentially represents the scale setting of QCD, which will be key topic of discussion later in this paper.
For three flavors, the latest value found is $C\sim 1.274 \times 10^{-11}$ for $\Lambda = 440 \ \text{MeV}$~\cite{Bae:2008ue,Wantz:2009it}.
However, this value can vary by almost an order of magnitude if $\Lambda$ is varied by 15\%. 

The interacting instanton liquid model (IILM) is a more sophisticated model that accounts for instanton-instanton interactions and was applied to the problem at hand in Ref.~\cite{Schafer:1996wv}.
 To solve for the free energy in this model, grand canonical Monte-Carlo simulations of the partition function are required.
The results can be fit to the form
\begin{equation}
F_{IILM}(\theta,T) = -D(T)\cos(\theta),
\end{equation}
with
\begin{equation}
D(T) \simeq \frac{e^{d_0}\Lambda^4}{(T/\Lambda)^{-d_1}}\exp\left[d_2 \left(\ln \frac{T}{\Lambda}\right)^2 + d_3 \left(\ln \frac{T}{\Lambda}\right)^3 + \cdots \right],
\end{equation}
where the values for $d_i$ are given in Ref.~\cite{Wantz:2009it} as a piecewise function of temperatures at the flavor mass thresholds (coincidentally, values for $d_0$ and $d_1$ do not differ from the analogous terms calculated in Ref.~\cite{Bae:2008ue}).
The authors point out that the largest uncertainties arise from the scale setting of using $\Lambda=400 \ \text{MeV}$ and emphasize at various stages that the IILM should be compared with lattice QCD results.

Since lattice QCD calculations are numerical and ultimately yield dimensionless numbers, lattice scale setting is of vital importance to relate to lattice results to reality.
The way this is typically done is to choose an observable (preferably one that can be calculated to high precision with little to no unphysical lattice artifacts from volume or finite lattice spacing) and match it onto some measured/derived quantity from experiment.
This is often done with matching onto heavy-quark potentials~\cite{Sommer:1993ce, Luscher:2009eq, Luscher:2010iy}, but this can also be done for spectroscopy of confinement scale masses, such as the omega baryon~\cite{Lin:2008pr}.
For our pure-glue calculation, the most reliable scale setting is to use the string tension, $\sigma$, which can be used to define the critical temperature $T_c$ in physical units.
Once this is done, all scales can be expressed in units of the critical temperature $T_c$ which makes it natural to fit to the forms
\begin{equation}\label{eq:Tc_forms}
\frac{CT_c^4}{(T/T_c)^n}\quad,\quad \frac{e^{d_0}T_c^4}{(T/T_c)^{-d_1}}\exp\left[d_2 \left(\ln \frac{T}{T_c}\right)^2 + d_3 \left(\ln \frac{T}{T_c}\right)^3 \right] \ .
\end{equation}
It should be noted that when using the string tension (or any other heavy-quark scale) for setting the scale, the critical temperature in Yang--Mills~\cite{Lucini:2005vg} is roughly twice that of full QCD~\cite{Borsanyi:2010bp,Bazavov:2011nk, Bhattacharya:2014ara}.
This is to be expected, as the dynamical quarks play a non-trivial role in the phase transition.

\section{Lattice simulations and scale setting}\label{sec:lattice}
We perform the lattice calculation of the free energy and its expansion in the $\theta$ angle using the Wilson plaquette action to discretize the SU(3) Yang--Mills continuum theory:
\begin{equation}
  \label{eq:plaquette-action}
  \cS_W \; = \; \beta \sum_{P} \left( 1 - \frac{1}{3} \mathrm{Re}\; \tr [ U_P ] \right) \comma
\end{equation}
where $U_P$ is the ordered product of gauge links along the plaquette and $\beta$ is the lattice gauge coupling.
This lattice action is well known and, in particular, a great deal of data exists in the literature regarding the scale setting procedure of the corresponding lattice system.
We will make use of this information when setting the temperature scale of our lattice simulations.
The lattice action in Eq.~\eqref{eq:plaquette-action} contains only one free parameter, $\beta$, which sets the lattice spacing through dimensional transmutation.
Once the lattice spacing is fixed, one can give it physical units by measuring a physical observable and relating it to an experimental value.

We are interested in spanning a large interval of temperatures with our simulations.
In order to accomplish that we fix the number of lattice sites in the temporal direction $\Nt$ to 6 and set the physical temperature
\begin{equation}
  \label{eq:lattice-temperature}
  T\left(	\beta,\Nt	\right) \; = \; \frac{1}{a(\beta) \Nt}
\end{equation}
by tuning a value for the lattice coupling constant $\beta$.
In Eq.~\eqref{eq:lattice-temperature} we explicitly show the dependence of the lattice spacing $a$ on $\beta$.
Working at a fixed $\Nt$, the lattice spacing alone determines the temperature. This approach is widely used in the literature~\cite{Borsanyi:2012ve}.
To check for possible lattice discretization effects in our results and establish a continuum limit, we simulate the same temperature at a smaller lattice spacing by simply increasing the temporal lattice sites (we use $\Nt = 8$).
Once the lattice spacing is fixed, the physical spatial volume is determined by the number of lattice points $\Ns$ in the three spacial dimensions: $L^3 = (a \Ns)^3$.
One complication of our approach is that the lattice spacing gets smaller at higher temperatures, implying smaller spacial volumes at fixed aspect ratio $\Ns / \Nt$.
As a consequence it is of paramount importance to check for possible finite-volume effects at high temperatures.

\begin{figure}[ht]
	\centering
		\includegraphics[width=.9\textwidth]{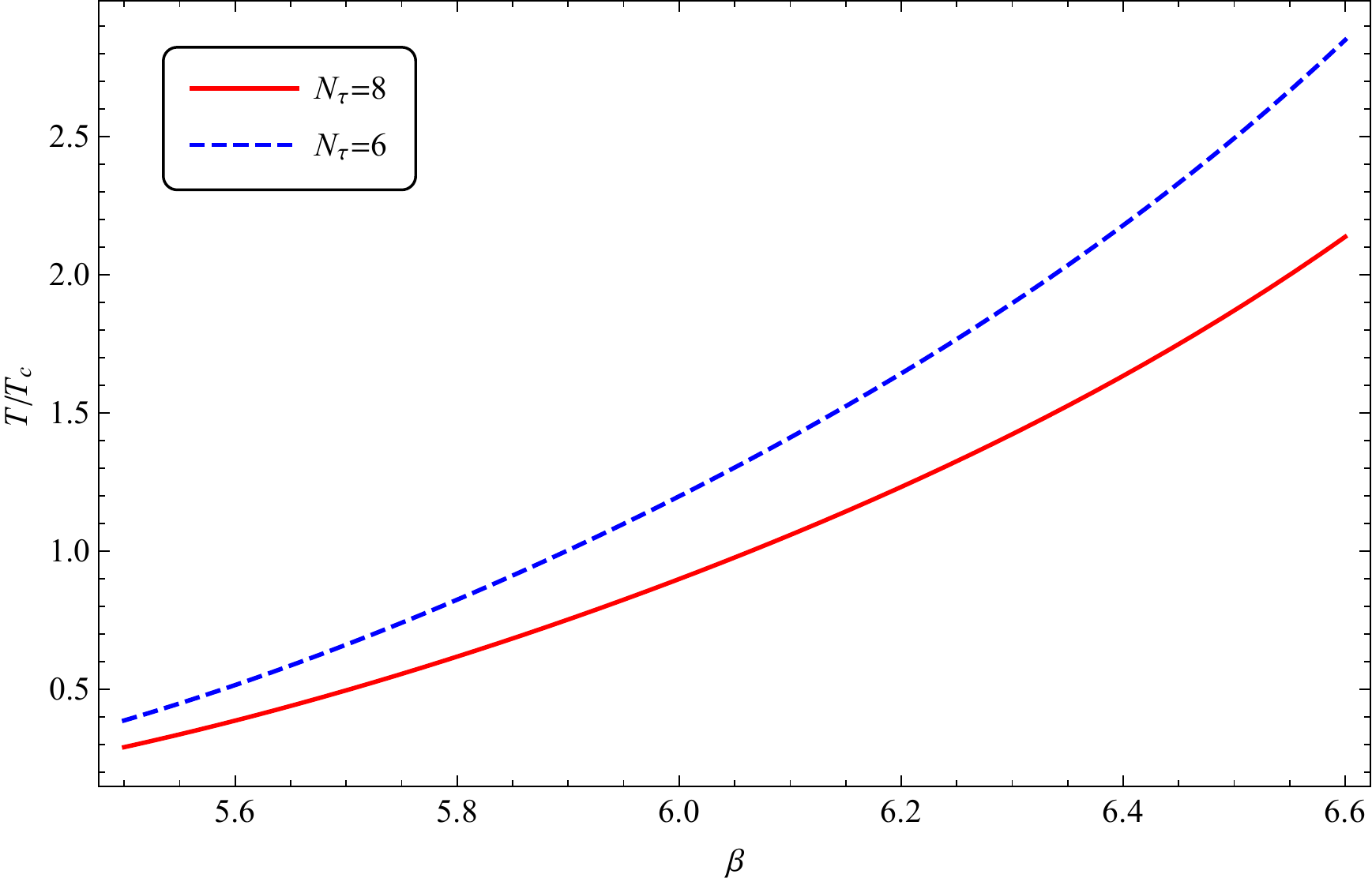}
	\caption{Temperature as a function of $\beta$, given $\beta_c = 6.338$ and $\Nt = 12$.}
	\label{fig:figures_TbyTc(beta)}
\end{figure}

Following Eq.~\eqref{eq:lattice-temperature}, setting the physical temperature scale has been traded for deducing the lattice spacing as a function of the bare lattice parameters.
In order to do so, one can follow a variety of approaches that usually involve measuring a physical quantity with the dimensions of a mass or a length in a zero--temperature setup.
We adopt the strategy of using the string tension $\sigma$, which is well understood for a Yang--Mills theory:
\begin{equation}\label{eq:TbyTcLucini}
	\frac{T}{T_c}
	\; = \;
	\frac{a\sqrt{\sigma}(\beta_c)\Nt_c}{a\sqrt{\sigma}(\beta)\Nt}
\end{equation}
where $\beta_c$ gives the critical temperature $T_c$ in a box of thermal extent $\Nt_c$ from the temporal Polyakov loop susceptibility.
In particular we choose the most continuum-like point at which the thermal phase transition has been studied for SU(3) with Wilson plaquette action~\cite{Lucini:2004my} ($\beta_c = 6.338$, $\Nt_c = 12$) and we interpolate numerical results for $a\sqrt{\sigma}$ as a function of $\beta$ from Ref.~\cite{Lucini:2004my}.
The interpolation uses the same function described in Ref.~\cite{Lucini:2005vg} and later refined in Ref.~\cite{Lucini:2012wq}.
In Fig.~\ref{fig:figures_TbyTc(beta)} we show $T/T_c(\beta)$ for two temporal extents $\Nt =6,\, 8$.

We can compare this approach with scale setting methods that use a different physical quantity, like the Sommer radius $r_0$, and quantify a systematic error for our procedure.
We have verified that setting the scale via both the static quark potential quantity $r_0$, as described in Ref.~\cite{Necco:2001xg}, and the continuum-extrapolated ratio of the critical temperature and the string tension $T_c/\sqrt{\sigma}$ as described in Ref.~\cite{Lucini:2004my} give $T/T_c$ values which agree with our method up to 1\% -- 1.5\% corrections in the temperature range explored in the paper.
Across the rest of the paper we report either results in units of the lattice spacing or results in units of the critical temperature.
This is to avoid giving physical units to $a$ or $T_c$; for a SU(3) Yang--Mills theory like the one we simulate, the systematic error associated to giving physical units to the lattice quantities is of the 4\% -- 7\% level~\cite{Durr:2006ky}.

Our gauge configurations are generated using the \emph{Chroma} software system~\cite{Edwards:2004sx} with GPU acceleration supplied by QDP-JIT/PTX~\cite{Winter:2014npa}.
The update algorithm is a standard Cabibbo--Marinari heatbath~\cite{Cabibbo:1982zn} for SU(3), alternated with 8 steps of standard over-relaxation to reduce correlations among subsequent configurations.
For each temperature $T/T_c$ we simulate two complementary streams of configurations: one starts from a random gauge configuration, while the other is initialized to a configuration where each link is a unit matrix.
On both streams we discard the first 10000 updates and monitor basic local observables like the spatial and temporal plaquettes and the Polyakov loops in the four dimensions.
For all temperatures the two streams thermalize to a common value of plaquette and Polyakov loop; this allows us to effectively double our statistics.
More details on the topological charge and topological susceptibility analysis are given in the next section. Our full set of ensembles and measurements is shown in Table~\ref{tab:datatable}.

\section{Topological susceptibility on the lattice}\label{sec:analysis}


\begin{table}
\begin{tabular}{|l|llrrr|ll|ll|ll|ll|}
\hline
$T/T_c$ & $\beta$ 	&   $\ass$	 	& $\Nt$ & $\Ns$ & $N_{meas}$&	\multicolumn{8}{|c|}{central value $\chi^{1/4}/T_c\pm\delta\chi^{1/4}/T_c$ statistical error for}												\\
	 	&		 	&   		 	& 		&		& 			&	\multicolumn{2}{c}{$\chi_\Real$}	&	\multicolumn{2}{c}{$\chi_\Rounded$}	&	\multicolumn{2}{c}{$\chi_\Artifact$}	&	\multicolumn{2}{c|}{$\chi_\Fit$}		\\
\hline\hline
 1.2 & 6.001 & 0.2161 & 6 &  64 & 14000 & 0.3880 & 0.0012 & 0.3814 & 0.0012 & 0.3871 & 0.0012 & \best{0.4192} & \best{0.0013}\\
1.31 & 6.053 & 0.1979 & 6 &  48 & 15600 & 0.3495 & 0.0009 & 0.3130 & 0.0009 & 0.3392 & 0.0010 &       0.3691  &       0.0011 \\
     &         &        &   &  64 & 36000 & 0.3424 & 0.0006 & 0.3358 & 0.0006 & 0.3402 & 0.0007 &       0.3703  &       0.0007 \\
     &         &        &   &  80 & 14000 & 0.3426 & 0.0010 & 0.3389 & 0.0010 & 0.3416 & 0.0010 & \best{0.3735} & \best{0.0011}\\
     &  6.242 & 0.1484 & 8 &  64 & 33998 & 0.3634 & 0.0010 & 0.3493 & 0.0010 & 0.3520 & 0.0010 &       0.3687  &       0.0010 \\
     &         &        &   &  96 & 14000 & 0.3556 & 0.0015 & 0.3533 & 0.0014 & 0.3537 & 0.0015 &       0.3703  &       0.0015 \\
 1.4 &  6.095 & 0.1852 & 6 &  64 & 54000 & 0.3153 & 0.0005 & 0.3077 & 0.0005 & 0.3095 & 0.0005 & \best{0.3370} & \best{0.0005}\\
 1.5 &  6.139 & 0.1729 & 6 &  64 & 54000 & 0.2928 & 0.0005 & 0.2833 & 0.0005 & 0.2814 & 0.0005 & \best{0.3068} & \best{0.0005}\\
 1.6 &  6.182 & 0.1621 & 6 &  64 & 53998 & 0.2721 & 0.0005 & 0.2587 & 0.0005 & 0.2568 & 0.0005 & \best{0.2799} & \best{0.0005}\\
 1.7 & 6.223 & 0.1525 & 6 &  64 & 24000 & 0.2536 & 0.0008 & 0.2330 & 0.0008 & 0.2369 & 0.0008 & \best{0.2585} & \best{0.0008}\\
 1.8 & 6.263 & 0.1441 & 6 &  64 & 24000 & 0.2343 & 0.0008 & 0.2005 & 0.0009 & 0.2178 & 0.0008 &       0.2368  &       0.0008 \\
     &         &        &   &  80 & 32000 & 0.2320 & 0.0006 & 0.2262 & 0.0006 & 0.2185 & 0.0006 & \best{0.2368} & \best{0.0006}\\
     & 6.471 & 0.1080 & 8 &  96 & 14000 & 0.2306 & 0.0016 & 0.2170 & 0.0017 & 0.2236 & 0.0015 &       0.2312  &       0.0016 \\
 1.9 & 6.301 & 0.1365 & 6 &  64 & 24000 & 0.2175 & 0.0009 & 0.1672 & 0.0011 & 0.2019 & 0.0008 &       0.2190  &       0.0009 \\
     &         &        &   &  80 & 34000 & 0.2164 & 0.0006 & 0.2095 & 0.0006 & 0.2026 & 0.0006 & \best{0.2189} & \best{0.0006}\\
1.99 &    6.550 & 0.0973 & 8 &  64 & 14795 & 0.2013 & 0.0034 & 0.1800 & 0.0036 & 0.1986 & 0.0029 &       0.2013  &       0.0034 \\
 2.0 &   6.338 & 0.1297 & 6 &  48 & 15600 & 0.2040 & 0.0018 & 0.1292 & 0.0027 & 0.1898 & 0.0016 &       0.2042  &       0.0018 \\
     &         &        &   &  64 & 25598 & 0.2032 & 0.0010 & 0.1390 & 0.0014 & 0.1893 & 0.0009 &       0.2041  &       0.0010 \\
     &         &        &   &  80 & 26000 & 0.2014 & 0.0008 & 0.1920 & 0.0008 & 0.1888 & 0.0007 &       0.2030  &       0.0008 \\
     &         &        &   &  96 & 14000 & 0.2004 & 0.0008 & 0.1961 & 0.0008 & 0.1900 & 0.0008 & \best{0.2038} & \best{0.0009}\\
 2.1 & 6.373 & 0.1235 & 6 &  80 & 24000 & 0.1880 & 0.0009 & 0.1749 & 0.0009 & 0.1774 & 0.0008 & \best{0.1889} & \best{0.0009}\\
 2.5 & 6.502 & 0.1037 & 6 & 128 & 14000 & 0.1497 & 0.0010 & 0.1479 & 0.0010 & 0.1494 & 0.0008 &       0.1492  &       0.0010 \\
     &         &        &   & 144 & 15797 & 0.1525 & 0.0008 & 0.1513 & 0.0008 & 0.1495 & 0.0006 & \best{0.1518} & \best{0.0008}\\
\hline
\end{tabular}
\caption{A summary of our lattice parameters and lattice results for the topological susceptibility in units of the critical temperature.
The temperature in units of the critical temperature and the corresponding lattice coupling $\beta$ are related by Eq.~\eqref{eq:TbyTcLucini}. We also report the corresponding value of the interpolated string tension. The combined number of measurements $N_{meas}$ on two different streams of configurations, one from a random start and one from a unit start.
Our representative values at each temperature are \best{emphasized}. They are all chosen from the globally-fit definition of $Q$ as given in Eq~\eqref{eq:Q_fit} because it leads to smaller discretization and finite-volume effects. These values are going to be used in all following analysis. The different definitions of the topological charge $Q$ from which the susceptibility is estimated are described in the text in Eq.~\eqref{eq:Q}.
}
\label{tab:datatable}
\end{table}

\begin{figure}[th]
	\includegraphics[width=0.95\textwidth]{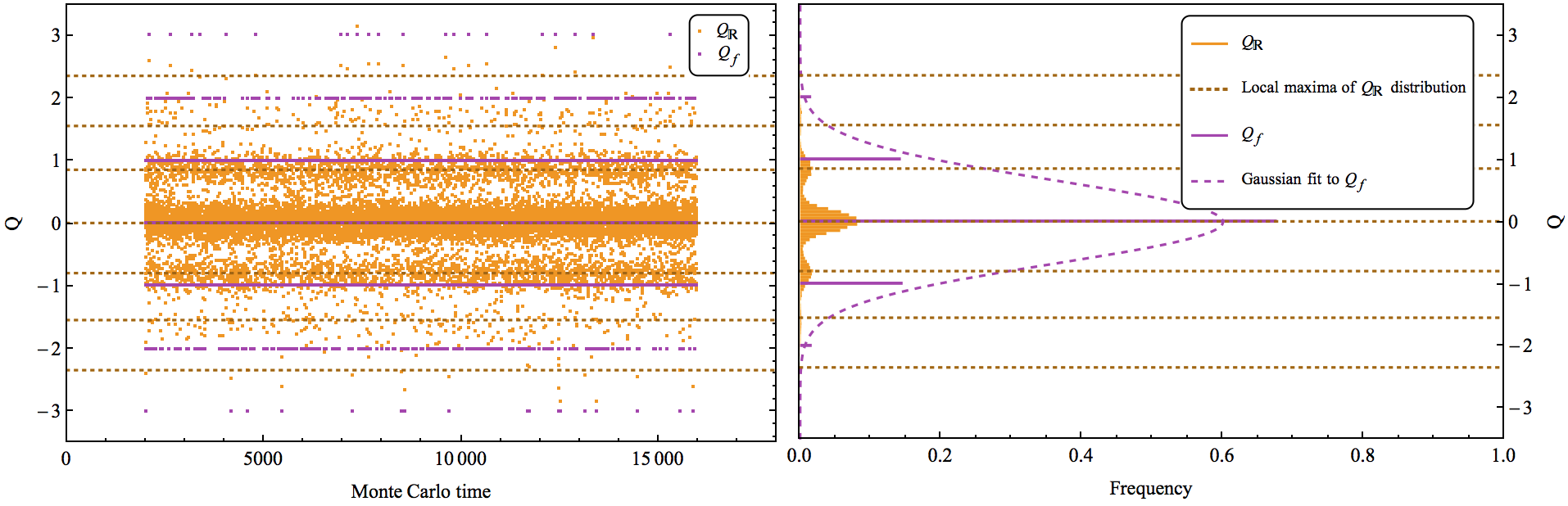}
	\caption{The topological charge as a function of Monte Carlo time for the 80$^3\times$6 $\beta=6.301$ ($T/T_c = 1.90$) ensemble.
	In the left panel, the yellow points are the measurements of $Q_\Real$ and the purple points are the globally-fit values $Q_\Fit$.
	In both panels, the brown dotted lines indicate the local maxima for the distribution of $Q_\Real$.
	The right panel shows a histogram of those points with a bin width of 0.05.
	The dashed purple line is a gaussian distribution with a standard deviation given by the second moment of the $Q_\Fit$ distribution.
}\label{fig:distribution-compare}
\end{figure}

We measure the topological charge on the lattice using the bosonic field-theoretical definition that is built upon discretizing the continuum formula on the right of Eq.~\eqref{eq:theta_term_topology}:
\begin{equation}
  \label{eq:lattice-topological-charge}
  Q_L \; \equiv \; \frac{1}{32\pi^2}\sum_x \epsilon_{\mu \nu \rho \sigma} \textrm{Tr}[U_{\mu \nu}(x) U_{\rho \sigma}(x)] \ ,
\end{equation}
where $U_{\mu \nu}(x)$ is the $\mu\nu$-plane plaquette at the lattice point $x$.
The above definition of lattice topological charge converges to the continuum definition when the lattice spacing is sent to zero, but it relies on the fact that the gauge fields are smooth.
The ultraviolet fluctuations of the gauge configurations are smoothed out via the cooling method~\cite{Smith:1998wt} before $Q_L$ is measured.
The cooling method is well tested and understood in the context of topological charge measurements on the lattice and it is equivalent to a smearing procedure or a gradient flow smoothing~\cite{Bonati:2014tqa}.
Empirically it was noted that this smoothing procedure is such that the multiplicative and an additive renormalization constants to $Q_L$ are close to one and zero, respectively.
This was understood theoretically in Ref.~\cite{Bonati:2014tqa}.
Our use of the cooling method should be viewed as a cheaper alternative to the gradient flow method which will be employed in our analysis when going beyond the Yang--Mills case.

Although UV fluctuations are removed by the cooling procedure while keeping topological properties largely unchanged, lattice artifacts affect $Q_L$.
Since we are interested in continuum physics, we use four different definitions of the topological charge based on $Q_L$.
The following definitions should all agree in the continuum limit and their behavior as a function of $a$ will help us characterize lattice discretization effects on our final results:
\begin{subequations}\label{eq:Q}
\begin{align}
	Q_{\Real}		&=	Q_L	\label{eq:Q_real}		\\
	Q_{\Rounded}	&=	\textrm{round}(Q_L)	\label{eq:Q_rounded}	\\
	Q_{\Artifact}	&=	\textrm{round}(Q_L)-\textrm{``narrow instantons''}		\label{eq:Q_artifact}	\\
	Q_{\Fit}		&=	\textrm{round}(\alpha Q_L) \quad \quad \alpha=\textrm{min}\left<[\alpha Q_L - \textrm{round}(\alpha Q_L)]^2\right>                           \label{eq:Q_fit}
\end{align}
\end{subequations}
where the \Real, \Rounded, \Artifact, and \Fit\ subscripts indicate the real-valued lattice definition of Eq.~\eqref{eq:lattice-topological-charge}, the rounded definition, a lattice-artifact corrected definition that subtracts contributions from narrow instantons~\cite{Smith:1998wt}, and a globally-fit definition that redefines the charge based on the properties of the whole distribution~\cite{DelDebbio:2002xa,Durr:2006ky}.
We measure $Q_L$ every $10$ Monte-Carlo updates and this gives us an autocorrelation time of 1 measurement or smaller for all $\Nt=6$ lattices and between 2 and 3 measurements for all $\Nt=8$ ensembles.
In Fig.~\ref{fig:distribution-compare} we compare $Q_{\Real}$ and $Q_{\Fit}$ for $T/T_c=1.90$ ($\beta=6.301$).

We define the lattice topological susceptibility from each of the definitions above:
\begin{equation}
  \label{eq:lattice-topological-susceptibility}
  \chi_i \; \equiv \; \lim_{V \rightarrow \infty} \frac{\left< Q_i^2 \right >}{V} \ ,
\end{equation}
where the index $i$ runs over the definitions in Eq.~\eqref{eq:Q_real}-\eqref{eq:Q_fit} and $V=a^4(\Ns)^3\Nt$.
We define our infinite volume susceptibility by measuring $\chi$ on different spatial volumes and by choosing the value on the largest volume that does not show signs of change.
In practice we find that the definition $\chi_{\Artifact}$ has the largest discretization effects while $\chi_{\Fit}$ has negligible effects due to finite lattice spacing.
In the following section we estimate finite-volume and discretization effects for all value of $T/T_c$ explored.
This will allow us to choose a final value of $\chi$ at each $T/T_c$ that is free of artifacts and can be fitted to models in Sec.~\ref{sec:res}.
All results are summarized in Table~\ref{tab:datatable}.

\section{Discretization and finite-volume errors}\label{sec:syst}
\begin{figure}[th]
	\includegraphics[width=0.75\textwidth]{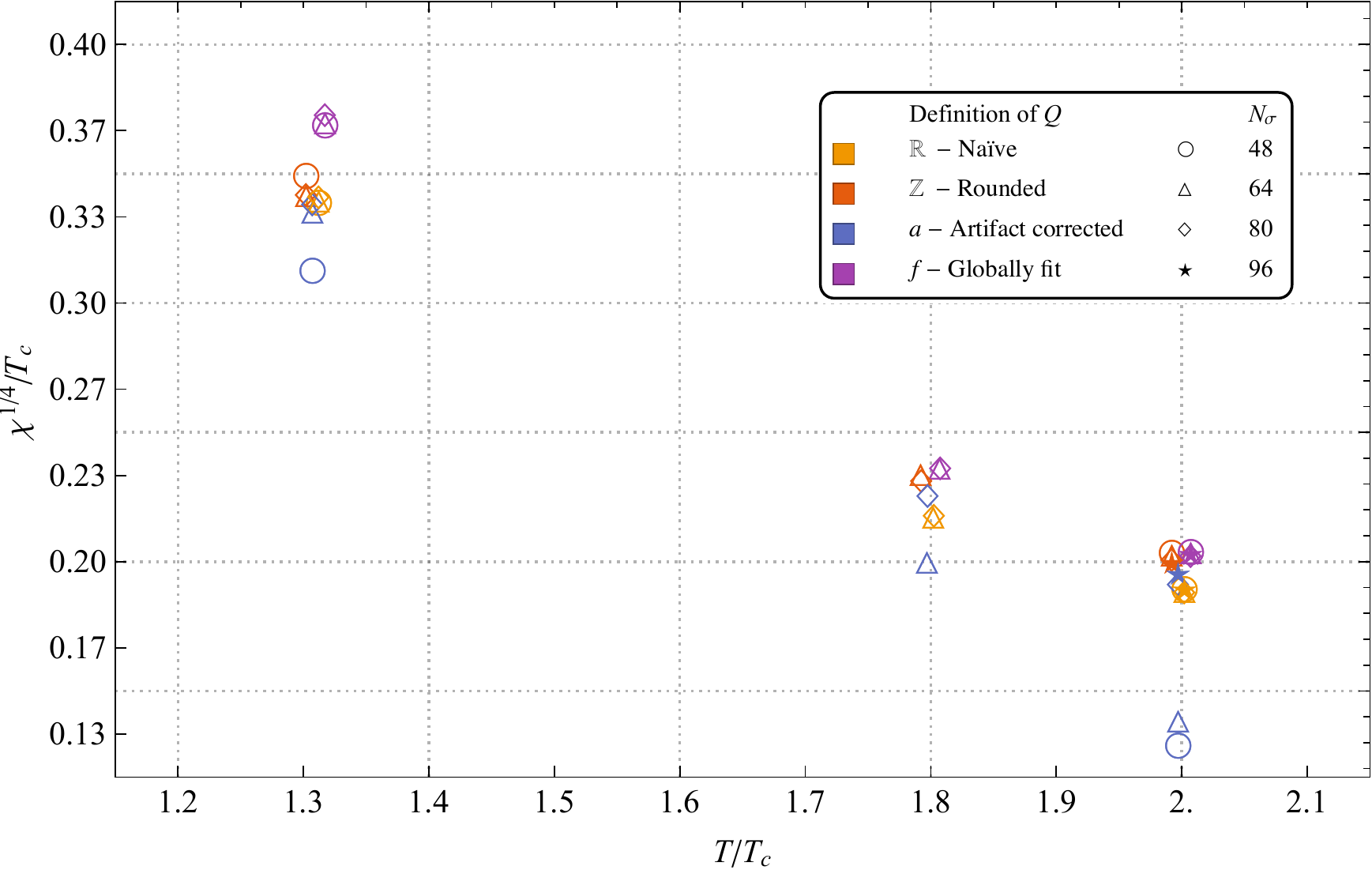}
	\caption{\label{fig:volume}
A study of finite-volume effects at three different temperatures.
All lattices have a temporal extent $\Nt=6$.
Each ensemble is slightly offset from its temperature for ease of visibility.
The statistical errors are smaller than the markers shown.
}
\end{figure}

\begin{figure}[th]
	\includegraphics[width=0.75\textwidth]{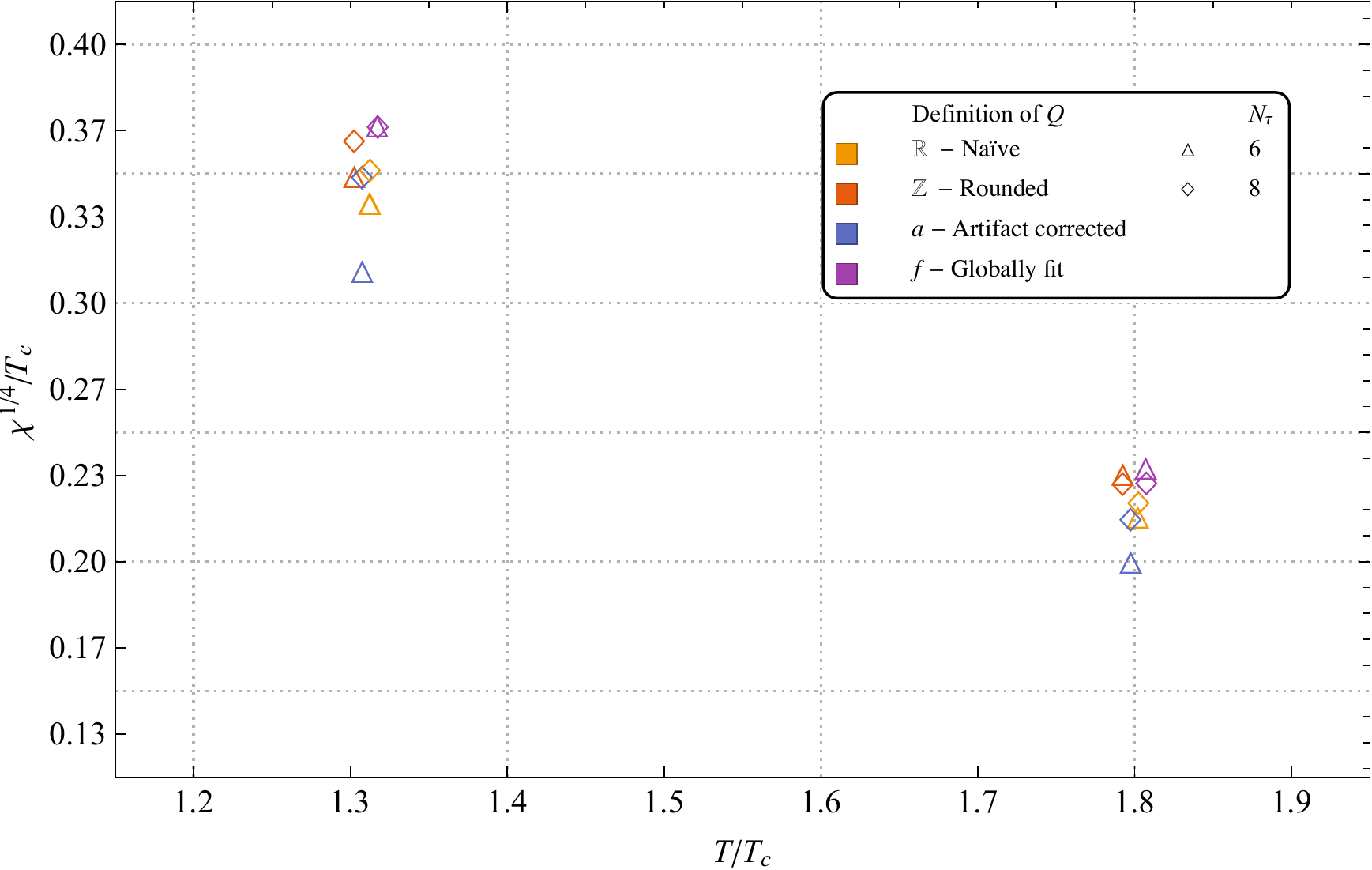}
	\caption{\label{fig:spacing} 
A study of lattice spacing effects at two different temperatures.  
At $T/T_c$=1.31 the lattices shown are $48^3\times6$ and $64^3\times8$ which have exactly the same physical volume, while at $T/T_c$=1.8 the lattice shown are $64^3\times6$ and $96^3\times8$, which have physical spacial extents that differ by only 13\%.
Each ensemble is slightly offset from its temperature for ease of visibility.
The statistical errors are smaller than the markers shown.
}
\end{figure}

In this section we study the systematic errors associated with having a discretized finite volume and a finite lattice spacing.
We start by looking at the various definitions $\chi_i$ at fixed $T/T_c$ and for increasingly larger boxes.
When the physical volume is large enough, we expect to obtain a constant topological susceptibility corresponding to Eq.~\eqref{eq:lattice-topological-susceptibility}.
For the simple rounded definition $\chi_{\Rounded}$ (shown in red in Fig.~\ref{fig:volume}), the smaller volumes have a consistently higher topological susceptibility.
In contrast, the artifact-corrected integer definition $\chi_{\Artifact}$ (shown in blue), has consistently smaller topological susceptibilities for smaller volumes.
The unrounded $\chi_{\Real}$ (yellow) and globally-fit rounded $\chi_{\Fit}$ (purple) definitions have essentially no finite-volume corrections, but the unrounded measure is systematically lower than the globally-fit definition.
At high temperature (small lattice spacing), $\chi_{\Fit}$ matches $\chi_{\Rounded}$ whereas $\chi_{\Real}$ does not.
For this reason we choose $\chi_{\Fit}$ as our most reliable definition.
This comports with the findings of Ref.~\cite{Durr:2006ky}.
Because of the concordance between the different volumes it seems as though many of the ensembles are all in the infinite-volume limit.
It is apparent from Fig.~\ref{fig:volume} that $\chi_{\Fit}$ has hardly any finite-volume effects.  

In Fig.~\ref{fig:spacing} we show discretization effects at two different temperatures.
At $T/T_c$=1.31 we show the topological susceptibility for the same physical volume at two different lattice spacings.
At $T/T_c$=1.8 we show the topological susceptibility for physical spatial extents that differ by roughly 13\%.
However, since we know from our finite-volume study shown in Fig.~\ref{fig:volume} that the volume makes essentially no difference at high temperatures (except for the artifact-corrected definition $\chi_{\Artifact}$), we can assume the difference in physical volume to be negligible compared to the discretization errors.
For both temperatures, $\chi_{\Fit}$ does not change when the lattice spacing is decreased by $\sim 30\%$.
For the higher temperature, which corresponds to an overall smaller lattice spacing, a similar effect is observed for $\chi_{\Rounded}$.
Because the globally-fit definition of the topological susceptibility is essentially independent of the volume and lattice spacing, we use that definition for our final ``curated'' data set, shown in bold in Table~\ref{tab:datatable}.
In the following we only consider a statistical uncertainty on the data points since the systematics effects discussed above are negligible.

\section{Comparison between lattice results and models}\label{sec:res}
\begin{figure}[th]
	\centering
		\includegraphics[width=0.75\textwidth]{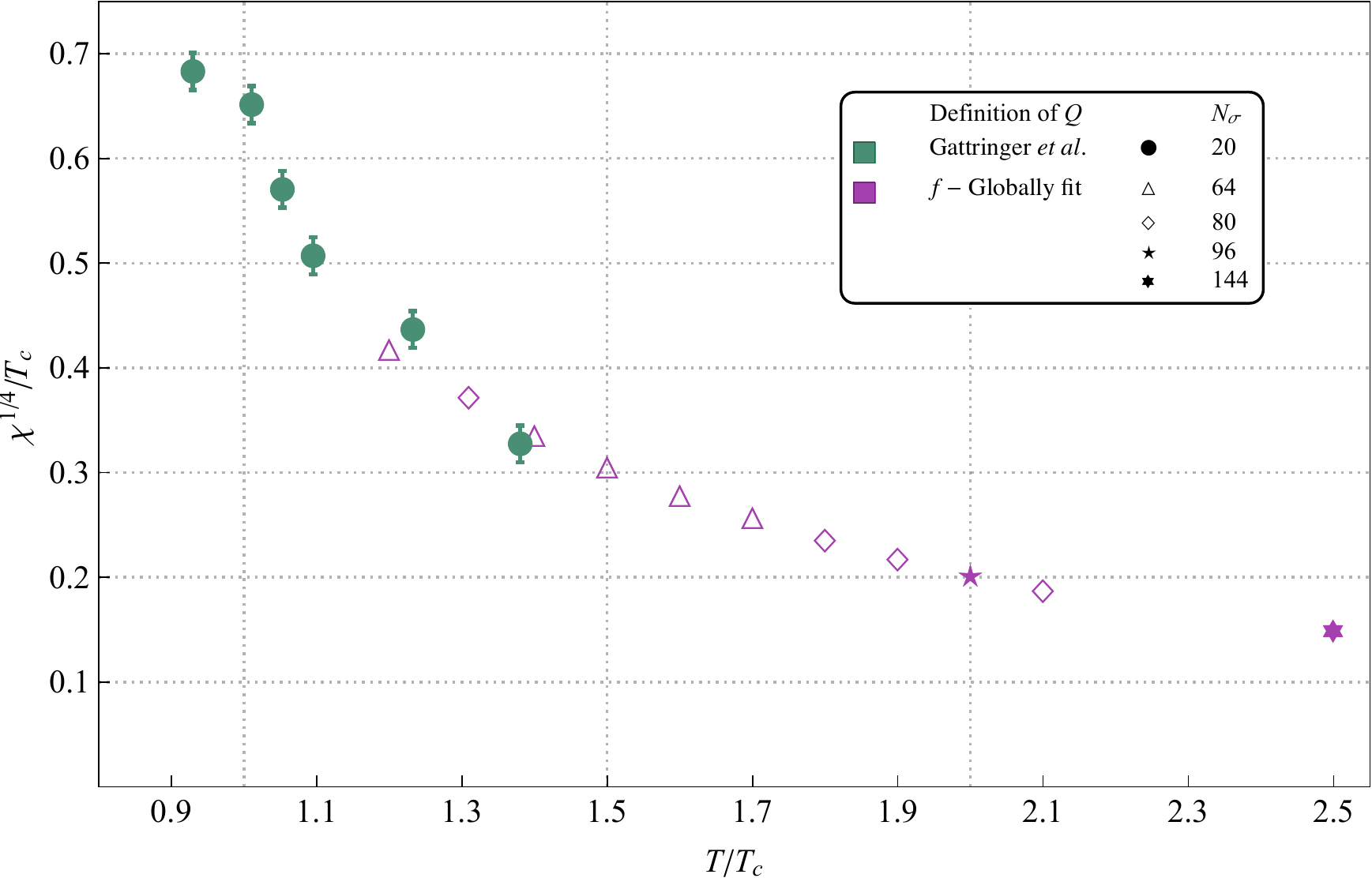}
	\caption{
		The largest volumes from Gattringer \emph{et al.}~\cite{Gattringer:2002mr} and our best $\Nt=6$ data points at each temperature.  
		The statistical errors on our points are smaller than the markers shown.
}
	\label{fig:best_data}
\end{figure}

We fit our best data points, those in bold in Table~\ref{tab:datatable} and shown in Fig.~\ref{fig:best_data}, to the forms shown in \eqref{eq:Tc_forms}.
Since the lattice calculations naturally yield $\chi/T_c^4$ and our temperatures are naturally in units of $T_c$ we actually fit
\begin{equation}
	\frac{\chi_{\text{DIGM}}}{T_c^4} 
	=
	\frac{C}{(T/T_c)^n}
	\quad,\quad
	\frac{\chi_{\text{IILM}}}{T_c^4} 
	=
	\frac{e^{d_0}}{(T/T_c)^{-d_1}}\exp\left[d_2 \left(\ln \frac{T}{T_c}\right)^2 + d_3  \left(\ln \frac{T}{T_c}\right)^3 \right]
\end{equation}
as functions of $T/T_c$ rather than $T$ itself.
This allows us to postpone consideration of the systematic error arising from setting the scale in physical units with an uncertain critical temperature.

\begin{figure}[th]
	\centering
		\includegraphics[width=0.75\textwidth]{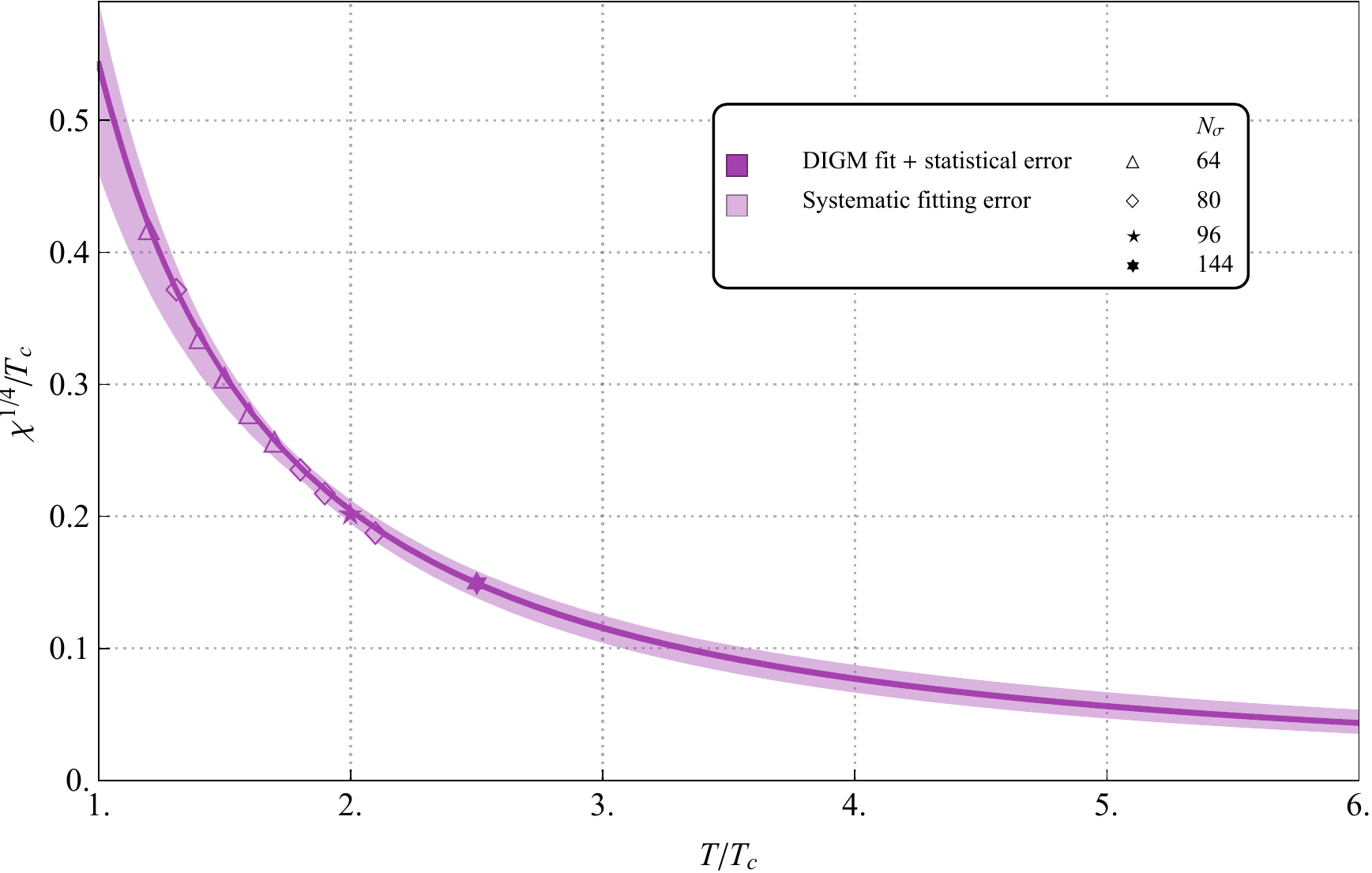}
	\caption{
		A fit of our best points to the dilute instanton gas model (DIGM).
}
	\label{fig:DIGM}
\end{figure}

\begin{figure}[th]
	\centering
		\includegraphics[width=0.75\textwidth]{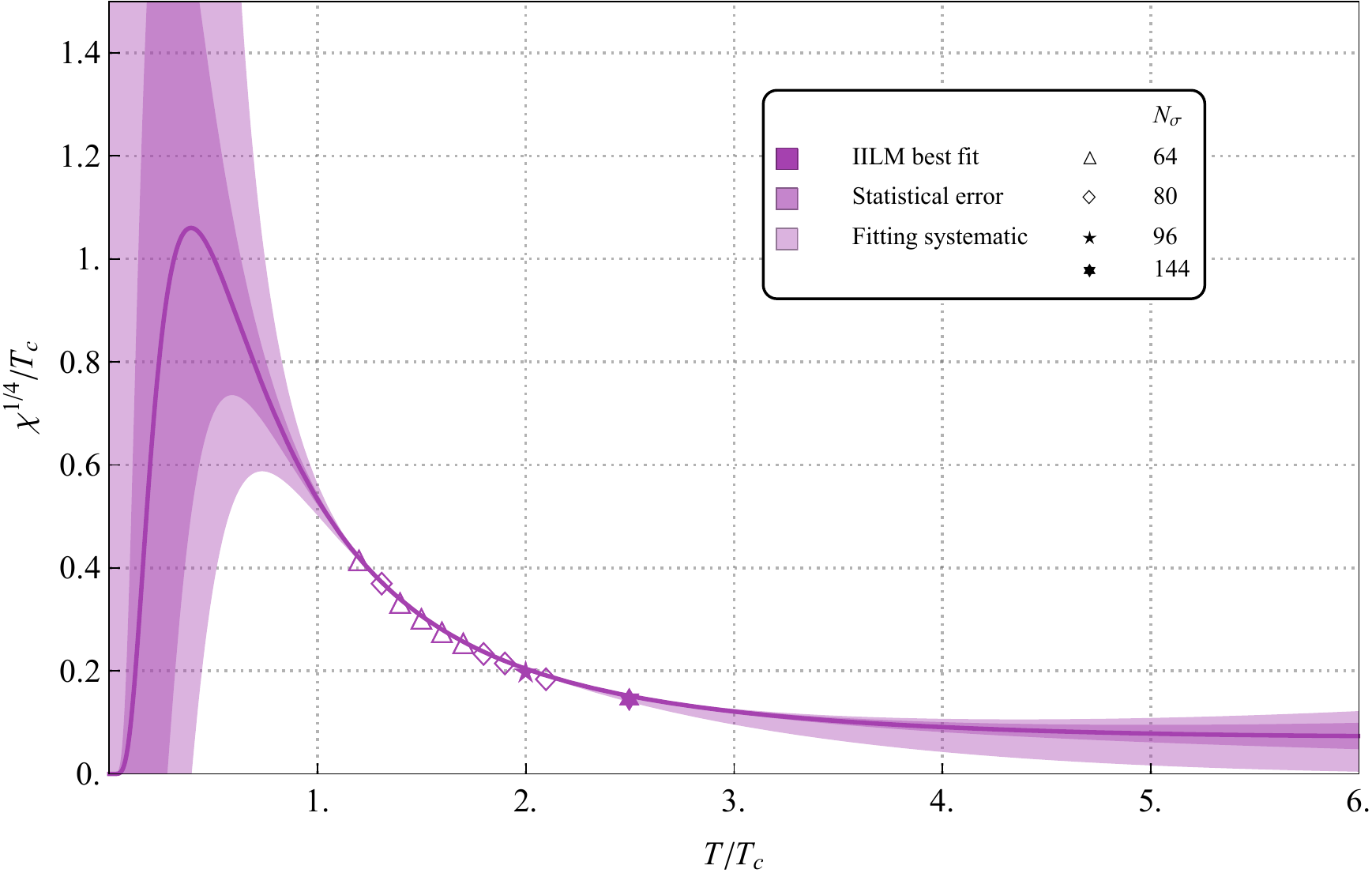}
	\caption{
		A fit of our best points to the interacting instanton liquid model (IILM).
}
	\label{fig:IILM}
\end{figure}

The central values shown in Fig.~\ref{fig:DIGM}~and~\ref{fig:IILM} are the results from fitting all the data points.
The fit parameters are shown in Table~\ref{tab:fits}, with a statistical uncertainty of one standard deviation (1$\sigma$).

\begin{table}[ht]
	\begin{center}
	\begin{tabular}{|lrl|}
		\hline
		\multicolumn{3}{|c|}{DIGM}	\\
		\multicolumn{3}{|c|}{$\chi^2$/d.o.f. = 1.2} \\
		\hline
		$C$	&	0.0869	&	$\pm$	0.0015	\\
		$n$	&	5.64	&	$\pm$	0.04	\\
		\hline
		\multicolumn{3}{c}{}\\
		\multicolumn{3}{c}{}\\
	\end{tabular}
	\hspace{2em}
	\begin{tabular}{|lrl|}
		\hline
		\multicolumn{3}{|c|}{IILM}		\\
		\multicolumn{3}{|c|}{$\chi^2$/d.o.f. = 1.7} \\
		\hline
		$e^{d_0}$	&	0.079	&	$\pm$	0.006	\\
		$d_1$		&	-4.9	&	$\pm$	0.5		\\
		$d_2$		&	-1.7	&	$\pm$	1.0		\\
		$d_3$		&	1.2		&	$\pm$	0.7		\\
		\hline
	\end{tabular}
	\end{center}
	\caption{Fit parameters for the DIGM and IILM fit to all of our best data points.}
	\label{tab:fits}
\end{table}

For the systematic fitting error on the DIGM we fit the same form to every subset of the best points with cardinality 3 or greater and take the outer envelope of all such fits' statistical 1$\sigma$ error bands.
This procedure is extremely conservative.
The resulting systematic fitting error is shown as a light purple band in Fig.~\ref{fig:DIGM}, where the statistical error band is entirely covered by the width of the line.

The DIGM is not reliable at low temperatures, and so that region is not shown in Fig.~\ref{fig:DIGM}.
The systematic errors at high temperature decrease---all the fits grow closer to the central value.

In the pure-glue case it is known~\cite{Gross:1980br} that the DIGM exhibits, at high temperatures, the leading behavior $\chi \sim T^{-n}$ where
\begin{equation}
	n = \frac{11}{3} N_c - 4 = 7
\end{equation}
for SU(3), compared to our numerical result $n=5.64\pm0.04$.
This suggests the temperatures we studied are not high enough to trust the high-temperature DIGM parameters.

The IILM is more reliable at low temperatures and unreliable at high temperatures.
Indeed, our best-fit IILM curve has a turning point at $T/T_c \approx 5.8$, beyond which it exhibits an unphysical increase of the topological susceptibility with temperature.

Because the IILM is not designed to capture high-temperature dynamics and has markedly more free parameters, the enveloping approach we used for estimating the systematic error bands for the DIGM is far too conservative and gets dominated by fits to a few high-temperature points.
Instead, we follow a jackknife-inspired procedure: for the central value we fit to all the best points, and for the systematic fitting error we fit to all subsets where one data point is omitted.
The outer envelope of all those fits' 1$\sigma$ statistical error bands is shown as the systematic fitting error in Fig.~\ref{fig:IILM}.
The inner, darker error band shown there is the statistical error band on the best fit of all of our data points.


\section{Illustrative example using lattice data to extract axion mass bounds }\label{sec:bounds}
As discussed in previous sections, the topological susceptibility $\chi$ for QCD as a function of temperature is required as input for the cosmological evolution of axions in the early universe.
In this section, we will take our lattice results for $\chi$ and carry out this procedure to calculate axion mass bounds with a significant disclaimer: our lattice calculation is affected by the absence of fermionic degrees of freedom.
This approach means that our numerical calculation was significantly cheaper than it would have been if we had calculated with full QCD, but can lead to a host of uncontrolled systematics.
Most obvious is the difference in critical temperatures between Yang--Mills and QCD\footnote{There is roughly a factor of two difference between QCD ($T_c \sim 154 \ \text{MeV}$~\cite{Borsanyi:2010bp,Bazavov:2011nk, Bhattacharya:2014ara}) and Yang--Mills ($T_c \sim 284 \ \text{MeV}$~\cite{Lucini:2005vg}).}.
The critical temperature $T_c$ is the primary scale where the theory deconfines.
It also indicates the onset of the fall of $\chi$ as temperature grows.  
In each of the individual steps to this point, this has not been a issue, as all quantities discussed in the result sections have been dimensionless ratios that do not depend on scale setting.
However, in the final steps required to extract a bound, the relative size of $T_c$ to the present day temperature of the cosmic microwave background, $T_\gamma$, plays a role.
It should be noted that this systematic will not be an issue for future calculations with physical QCD parameters.  For now, it prevents bounds relevant to reality from being extracted.

Let us describe the overall effect of setting the absolute scale $T_c$.
As discussed in Sec.~\ref{sec:misalignment}, the axion begins rolling down the potential at the temperature where $m_a\approx 3H$.
With simple algebra this may be rewritten
	\begin{equation}\label{eq:chi_H}
		\chi(T_1) \approx 9 f_a^2 H^2(T_1) \ .
	\end{equation}
The fortuitous form of $H$ in Eq.~\eqref{eq:H1Gev},
	\begin{equation}
		H^2(T) = \frac{\pi^2 }{90 M_{Pl}^2}\; g_*(T)\; T^4
	\end{equation}
makes it particularly nice to write the relation~\eqref{eq:chi_H} in terms of $T/T_c$, the temperature ratio that shows up in our lattice calculations.
One may show that the relation $m_a\approx 3H$ can be rewritten
\begin{equation}\label{eq:scale_setting}
	\frac{\chi}{T_c^4}(T_1/T_c)
	\; \approx \;
	\frac{\pi^2 f_a^2}{10 M_{Pl}^2}\; g_*(T_1/T_c \cdot T_c) \cdot \left(\frac{T_1}{T_c}\right)^4
\end{equation}
which allows us to put all of the absolute scale dependence into the argument of $g_*$.
Fortunately, at the temperatures of interest $g_*$ is insensitive to the difference between the critical temperature $T_c$ in QCD and the pure-glue theory~\cite{Wantz:2009it}---it is essentially a constant at and between those temperatures.
The particular temperature dependence of Eq.~\eqref{eq:scale_setting} also means that we do not need to translate our lattice results for $\frac{\chi}{T_c^4}$ into physical units, therefore avoiding what is usually the largest source of systematic uncertainties.

However, this does not imply that all the systematic errors are erased.
Instead, because $g_*$ captures the relevant degrees of freedom of reality, which includes quarks, rather than a quark-free universe, we still have an uncontrolled systematic error.
This systematic will be avoided when future lattice calculations use physical QCD parameters, while the uncertainty in the QCD critical temperature $T_c$ will be remain irrelevant thanks to the insensitivity of $g_*$.
Thus, we expect the reliability of future lattice calculations to be controlled entirely by the accuracy of the determination of $\chi/T_c^4$ and cosmological inputs, not the accuracy of the QCD critical temperature. 

Of course, with $\chi/T_c^4$ from QCD in hand, it will make sense to not simply estimate the onset of the axion field rolling when $m_a\sim3H$ but to instead solve the cosmological equations of motion numerically.
Moreover, it will be simple to put reliable bounds on the pre-inflation PQ-breaking scenario, constraining a combination of the initial $\theta$ parameter and the axion mass.
Therefore we emphasize our results in the rest of this section as the first step towards a full-fledged calculation of axion constraints from first principles.

\begin{figure}[t]
	\centering
	\includegraphics[width=0.75\textwidth]{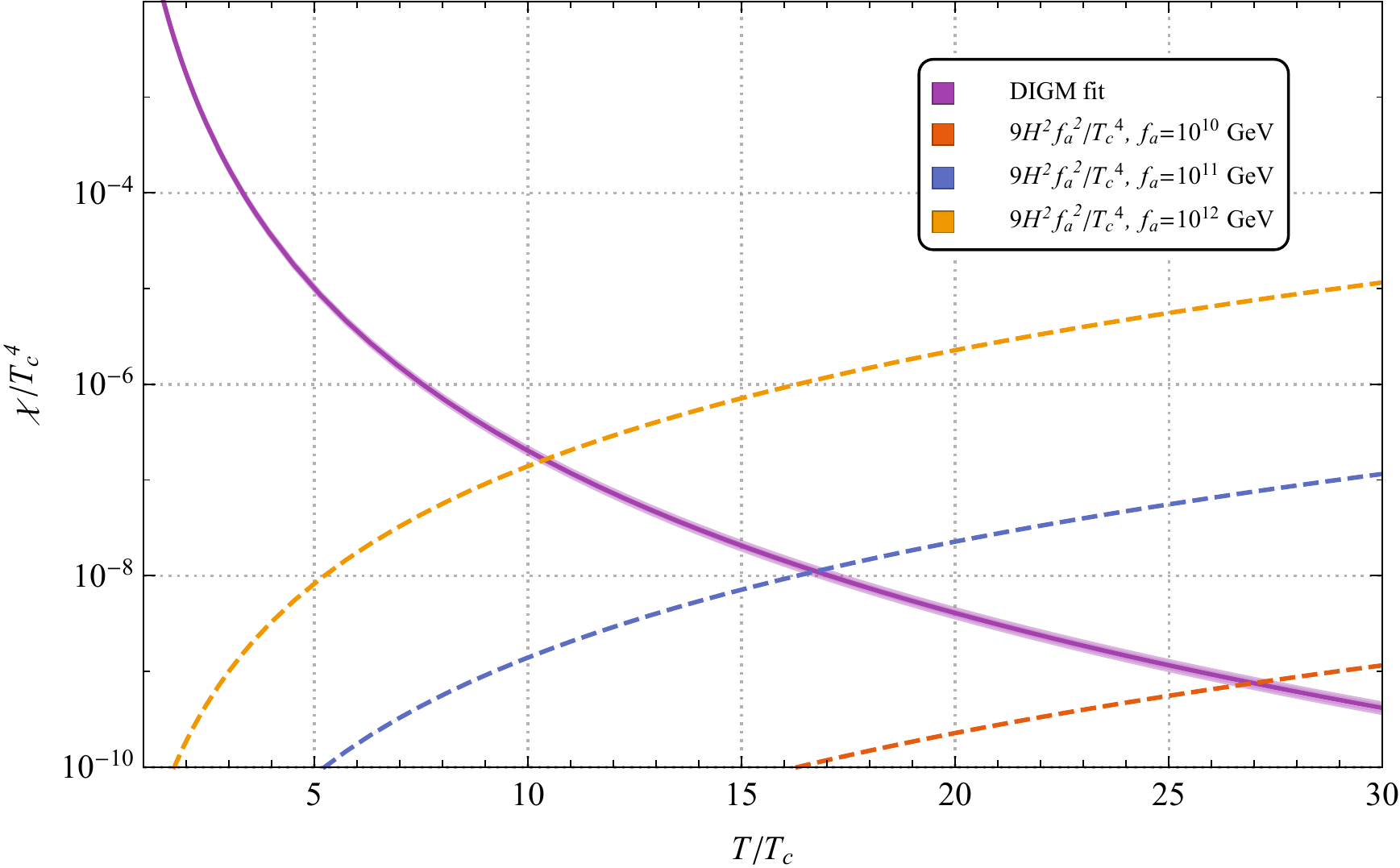}
	\caption{The DIGM-inspired fit with systematic fitting uncertainty shown with lines of $(3H)^2$ for three representative choices of $f_a$.}
	\label{fig:DIGM_Hsq}
\end{figure}

\begin{figure}[th]
	\centering
	\includegraphics[width=0.75\textwidth]{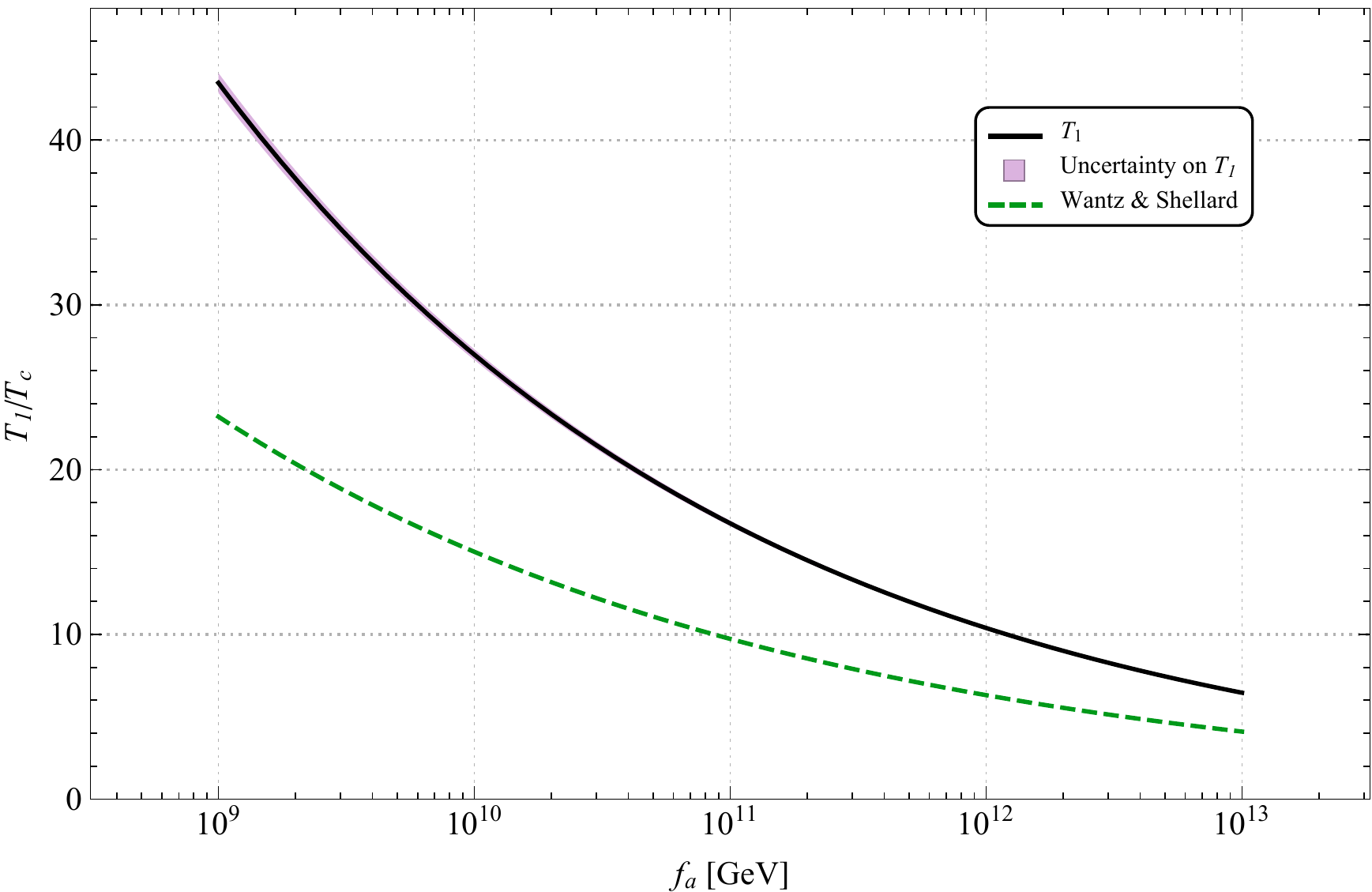}
	\caption{
	The value of $T_1$, the temperature where $\chi=9H^2f_a^2$, as a function of $f_a$ if $\chi$ is given by our DIGM-inspired fits.
	The purple band includes the systematic fitting uncertainty and the mild scale setting ambiguity added in quadrature.
        The dashed line is the results from Ref.~\cite{Wantz:2009it} when $T_c=154\ \text{MeV}$.
	}
	\label{fig:T1}
\end{figure}

In Fig.~\ref{fig:DIGM_Hsq} we plot our numerically-fit DIGM extrapolation of $\chi/T_c^4$, the left-hand side of Eq.~\eqref{eq:scale_setting} and three example right-hand sides for different choices of $f_a$ as a function of temperature.
The intersection gives $T_1/T_c$ as a function of $f_a$, which is shown in Fig.~\ref{fig:T1}.
In terms of the DIGM fit parameters, we have
\begin{equation}
\frac{T_1}{T_c} = \left[\frac{10 C}{\pi^2 g_*(T_1)}\left(\frac{M_{Pl}}{f_a}\right)^2\right]^{\frac{1}{4+n}}. 
\end{equation}
where the insensitivity of $g_*$ to temperatures in the regime of interest allows us to solve self-consistently with ease.
In a full calculation without extrapolation using a model, $T_1$ can be determined as a function of $f_a$ numerically from Eq.~\eqref{eq:scale_setting}.  In Fig.~\ref{fig:T1}, our extracted $T_1/T_c$ is compared with that of the IILM \cite{Wantz:2009it}.

With $T_1$ in hand, we can calculate the axion energy density using
\begin{equation}\label{eq:omega-a}
\Omega_a = \frac{\rho_a(T_\gamma)}{\rho_c} \quad,\quad \rho_c = 3.978 \times 10^{-11} \left(\frac{h^2}{0.701}\right)\ \text{eV}^4. 
\end{equation}
The resulting energy density from the misalignment mechanism is given by
\begin{equation}\label{eq:omega-a_eval}
h^2\Omega_a = 0.107(1)\;  \left(\frac{F(\theta_1) \theta_1^2}{2\gamma}\right)\left(\frac{f_a}{10^{12}\ \text{GeV}}\right)^{1.2074(8)}
\end{equation}
where the errors are determined by bootstrapping the fit results for the DIGM shown in Table~\ref{tab:fits}.
A bound can be derived for the post-inflation PQ-breaking scenario by starting from this density, setting $\langle \theta_1^2 \rangle = \pi^2/3$,  and comparing with the latest dark matter density from PLANCK~\cite{Ade:2013zuv}
\begin{equation}\label{eq:omega-dm}
\Omega_a h^2 \leq \Omega_{DM} h^2 = 0.1199 \pm 0.0027.
 \end{equation}

This procedure leads to an upper bound on the value of $f_a$ and Eq.~\eqref{eq:axion_mass_today} can be used to translate this into a lower bound on the axion mass today.
Putting all these pieces together using lattice inputs for Eq.~\eqref{eq:density}, we arrive at the bounds
\begin{equation}\label{eq:bound_results}
\parbox{0.5\textwidth}{	\raggedleft 
Post-inflation PQ breaking,\\  
physical $g_*$, $m_\pi$, $f_\pi$, $T_c$, \\ 
pure-glue $\chi$
}\quad 
	\begin{cases}
	f_a	& \leq	(4.10\pm0.04) \times 10^{11}\  \text{GeV} \\
	m_a & \geq 	(14.6\pm0.1) \ \mu \text{eV}
	\end{cases}
\end{equation}
which includes all the statistical, systematic and scale-setting errors from fitting to and extrapolating with the DIGM but does not include any cosmological uncertainties.   Also, to reiterate, this is only the bound from the misalignment mechanism and does not include string or domain wall decay, which would give even tighter bounds.
In comparison, the most comprehensive, up-to-date bound from the misalignment mechanism is given by $m_a \geq 21 \ \mu \text{eV}$ \cite{Wantz:2009it}.

Again, it should be noted that these results are for the topological susceptibility $\chi$ from a \textit{pure-glue} calculation and not from the full QCD theory with dynamical fermionic degrees of freedom.
That being said, this calculation is still illuminating and illustrates how uncertainty quantification due to strong dynamics can be propagated to the final steps of this calculation.

\section{Conclusions} \label{sec:concl}
We performed a lattice calculation of topological susceptibility for pure Yang--Mills theory at high temperatures and applied the results to the problem of axion cosmology.
There are two primary accomplishments of this work.

First, we developed the connection between actual lattice data and the well-known procedure for calculating the axion density, with controlled uncertainties in the input from the non-perturbative free energy.
Once lattice QCD results span the temperatures of interest, a simple interpolating fit can be used to solve the cosmological equations of motion.
In the mean time, we must rely on model-inspired fit forms extrapolate to the high temperatures of interest.
In this regard, we find that the DIGM-like fit, Eq.~\eqref{eq:Tc_forms}, works remarkably well for a theory consisting only of gluons, yielding a $\chi^2/\text{d.o.f} \sim 1.2$ with very precise lattice data (statistical errors roughly at the 0.2\% level).
However, we find that our points do not conform to the pure-glue expectation that the DIGM exponent $n$ is $7$, but that $n=5.64\pm0.04$.
Also, even when propagating the systematic errors, the final bounds were found to have total uncertainties only at the few percent level.
Thus, the DIGM-like fit proves to be a good continuous fit form for translating finite lattice calculations at discrete temperature values to an equation that can be used to solve the cosmology evolution equations numerically.
Moreover, the lattice can provide reliable errors on the fit parameters and these errors can be propagated to the final answers straightforwardly.

Second, we aimed to perform a detailed study of lattice systematic effects of lattice calculations of the topological susceptibility at high temperature.
To accomplish this, small statistical errors are a must.
Moreover multiple volumes, lattice spacings and discretized definitions of the topological susceptibility are required to assess systematic effects and add precision.
As expected, it was found that the lattice artifacts affect each of the discretized definitions of the topological susceptibility differently, and while lattice spacing effects were the predominant error in most of the definitions, one definition (artifact-corrected, $Q_a$) was extremely sensitive to finite volumes at high temperatures/small lattice spacing, leading to shifts of order 100\%.
Overall, it was found, as in Ref.~\cite{DelDebbio:2002xa,Durr:2006ky}, that the globally-fit definition, $Q_f$, had virtually no systematic errors and while a distinct systematic error could be seen in all of the other definitions, no effect larger than the statistical error bars could be quoted for this definition.

One disclaimer that has been pointed out frequently throughout this paper is that the lattice results were for a purely gluonic theory and the procedure for determining the axion mass bounds assumed full QCD.
This leads to multiple issues, which essentially reduce the results in this paper to simply be a preliminary step towards being useful in realistic experiments.
The low temperature axion mass depends on the pion mass and decay constant, which do not exist in a purely gluonic theory, and the relevant number of degrees of freedom, $g_*$, depend on the number of fermions below the temperatures of interest.
Second, QCD and pure Yang--Mills differ in deconfinement temperatures by roughly a factor of two when scales are fixed to heavy-quark scales (thus, correctly describing high temperature physics and heavy-quark physics cannot be done simultaneously).
Third, dynamical fermion zero-modes are expected to play an important role in the temperature behavior of the topological susceptibility, which are omitted in Yang--Mills lattice calculations.

Despite all of these limitations, it is tempting to see how the bounds behave compared to quoted values of the axion mass bounds.
The results in Eq.~\eqref{eq:bound_results} imply that the axion mass bound is weaker that the up-to-date value of  $m_a > 21 \ \mu\text{eV}$, and if true, would imply more viable parameter space for the post-inflation PQ-breaking scenario. Current and next generation ADMX experiments will throughly explore this additional axion mass parameter space over the next few years.
Moreover, these lattice-based bounds have a well defined and robust theoretical error associated with them, which is a very compelling feature compared to previous bounds.

While our bounds were derived without the inclusion of fermions in the lattice calculation, they are ample motivation for future studies and calculations in this direction.
From this work many of the concerns on volume limitations have been quelled, which suggests that lattice QCD calculations on smaller volumes may be sufficient and prohibitively large volumes may not be needed.
The lattice QCD issue of the increased auto-correlation and freezing of topological charge could still be significant, but one potential solution is to extract the topological susceptibility with fixed topology as in Ref.~\cite{Cossu:2013uua} or work with anisotropic lattices to boost the volume in space while keeping the temporal direction small to go to high temperatures.
Both of these approaches should be explored further at higher precision to accurately quantify the size of the systematic effects.
Once methods of extracting topological susceptibilities at high temperatures are reliable, the next step would be to perform this calculation for QCD.
Typically, calculations of this kind start with larger-than-physical quark masses (i.e. pion masses around $400$ MeV) with non-chiral fermion discretizations.
However, using lattices with physical quark masses and chiral domain-wall fermions~\cite{Bhattacharya:2014ara} is most likely next step given current computing resources and given that much of the involved scale setting procedure and non-perturbative tuning has been performed.
At that stage, the lattice QCD results will be directly applicable to cosmological simulations and will have direct connection with experimental searches.

\begin{acknowledgments}
We are indebted to David Kaplan for many useful discussions and pointing us towards these questions in axion cosmology. 
We would also like to thank Gianpaolo Carosi, Guido Cossu, Graham Kribs, Biagio Lucini, Pierre Sikivie, and Pavlos Vranas for useful discussions.
We generated gauge configurations using Chroma~\cite{Edwards:2004sx} on the GPU clusters (Edge and Surface) at LLNL.
Chroma was configured to use the QDP-JIT GPU library~\cite{Winter:2014npa} to accelerate the production.
This work was performed under the auspices of the U.S. Department of Energy by LLNL under Contract No.~DE-AC52-07NA27344.
M.I.B. is supported by DOE Grant No.~DE-FG02-00ER41132.
This research was partially supported by the LLNL Multiprogrammatic and Institutional Computing program through a Tier 1 Grand Challenge award.
\end{acknowledgments}

\bibliography{top_sus_axion} 

\end{document}